\documentclass{FrontiersinHarvard}

\usepackage{url,hyperref,lineno,microtype,subcaption}
\usepackage[onehalfspacing]{setspace}
\usepackage{multirow}   % <-- this enables \multirow
\usepackage{longtable}
\usepackage{array}
\usepackage{lineno}
\usepackage{microtype}

\providecommand{\doi}[1]{doi:\,#1}

\newcommand{\araa}{ARA\&A}%
\newcommand{\mnras}{MNRAS}%

\def\keyFont{\fontsize{8}{11}\helveticabold }
\def\firstAuthorLast{Mondal {et~al.}} %use et al only if is more than 1 author
\def\Authors{Indrajit Mondal\,$^{1}$, Prasanta Gorai\,$^{2,3,1*}$, Ankan Das\,$^{1}$, Suman Kumar Mondal\,$^{1}$, Rubén Fedriani\,$^{4}$, Xiaohu Li\,$^{5}$, Parama Mahapatra\,$^{1,6}$, Sabyaasachi Banik\,$^{1,7}$, Sheng-Li Qin\,$^{8}$ }
\def\Address{$^{1}$ \tiny Institute of Astronomy Space and Earth Sciences, P 177, CIT Road, Scheme 7m, Kolkata 700054, West Bengal, India\\
$^{2}$Rosseland Centre for Solar Physics, University of Oslo, PO Box 1029 Blindern, 0315, Oslo, Norway
$^{3}$ Institute of Theoretical Astrophysics, University of Oslo, PO Box 1029 Blindern, 0315, Oslo, Norway
$^4$ Instituto de Astrofísica de Andalucía, CSIC, Glorieta de la Astronomía s/n, 18008 Granada, Spain
$^{5}$Xinjiang Astronomical Observatory, Chinese Academy of Sciences, 150 Kexueyi St, Xinshi District, Ürümqi, Xinjiang 830011, China 
$^{6}$  Adamas University, Barbaria, Jagannathpur, District-24 Parganas (North), Kolkata 700126, West Bengal, India
$^{7}$  St. Xavier's College (Autonomous), 30, Mother Teresa Sarani, Kolkata 700016, West Bengal, India

$^8$ School of Physics and Astronomy, Yunnan University, Kunming 650106, China
}
\def\corrAuthor{Prasanta Gorai}

\def\corrEmail{prasanta.gorai@astro.uio.no }

\begin{document}
\onecolumn
\firstpage{1}

\title[ Molecular Lines Toward G10.47+0.03]{ Herschel/HIFI Observations of Molecular Lines Toward G10.47+0.03} 

\author[\firstAuthorLast ]{\Authors} %This field will be automatically populated
\address{} %This field will be automatically populated
\correspondance{} %This field will be automatically populated

\extraAuth{}% If there are more than 1 corresponding author, comment this line and uncomment the next one.
%\extraAuth{corresponding Author2 \\ Laboratory X2, Institute X2, Department X2, Organization X2, Street X2, City X2 , State XX2 (only USA, Canada and Australia), Zip Code2, X2 Country X2, email2@uni2.edu}

\maketitle
\begin{abstract}
 We present a spectral line analysis of the hot molecular core G10.47+0.03 (hereafter, G10). Our aim is to determine molecular abundances and excitation conditions across a wide spectral range inaccessible to ground-based observatories. We utilize archival data from the Herschel Space Observatory, obtained with the Heterodyne Instrument for the Far-Infrared (HIFI). We report here the detection of high-excitation CO, $^{13}$CO, and C$^{18}$O, H$_2$O isotopologues, HCO$^+$, HCN, HNC, CS, C$^{34}$S, SO, SO$_2$, H$_2$CS, and CH$_3$OH. CO, p-H$_2$O, CS, and HCN show similar velocity profiles with a narrow, blueshifted component, which may be linked to the outer outflow layer. Redshifted wings may indicate inner outflow activity. A Markov Chain Monte Carlo framework is employed to infer column densities and temperatures accurately. We also performed spectral energy distribution fitting to constrain the global physical parameters of G10, providing essential context for interpreting the molecular emission. The MCMC analysis revealed two excitation temperature components: a warm component ($\sim$30-65 K) and a hot component ($\sim$90-250 K). The higher temperatures indicate dense, hot gas typical of massive hot cores. The lower temperatures correspond to the warm, less dense envelope around the core. Transitions of H$_2$O, high-excitation CO, and HCN indicate outflowing gas and high-density shocked regions. These findings highlight G10's complex dynamical environment.
\end{abstract}
{\tiny
 \keyFont{Keywords:} astrochemistry, interstellar medium (ISM), Herschel/HIFI, molecular clouds, hot molecular core, star formation} %All article types: you may provide up to 8 keywords; at least 5 are mandatory.

\section{Introduction}
So far, about 340 molecular species have been detected in the interstellar medium (ISM) and circumstellar envelopes. Many of these species are complex and may play a role in the chemistry that could lead to the emergence of life \citep[see CDMS database\footnote{\url{https://cdms.astro.uni-koeln.de/classic/molecules}},][]{McGuire22}. 
Observations from ground-based telescopes have significantly advanced our understanding of molecular clouds and star-forming regions, enabling the discovery of new molecules and the mapping of their spatial distributions. However, a complete characterization of the chemical inventory in individual sources remains limited due to the lack of continuous, wide-frequency spectral line surveys. A large portion of this wide frequency is inaccessible from the ground due to atmospheric absorption.
Comprehensive spectral line surveys are essential for advancing our understanding of chemical complexity in the ISM, the processes driving star and planet formation, and the gas-phase/ice-phase pathways leading to prebiotic chemistry \citep{Garrod2013,das24,ligt25}. Such surveys provide access to a wide range of molecular and atomic transitions, enabling the determination of abundances, excitation conditions, and underlying physical parameters in diverse astrophysical environments. The development of high-resolution spectrometers, particularly aboard space-based observatories, have been instrumental in achieving these goals by extending spectral coverage beyond the limitations imposed by Earth’s atmosphere \citep{Zernickel12, Kama13, Kazmierczak14}.

The compact ($<$0.1 pc), hot ($\geq$100 K), and dense ($\geq$ 10$^6$ molecules per cubic centimeter) regions known as hot molecular cores (HMC) are exciting because they act as natural laboratories for the formation and evolution of complex organic molecules (COMs) \citep{dish98, mondal2023}. HMCs are chemically rich gas-phase and grain-surface chemistry conditions, allowing the formation of COMs \citep{bell19,jorg20}. These regions usually have a ``forest" of molecular lines, suggesting a wide range of chemical species, from basic molecules like water (H$_{2}$O) and methanol (CH$_{3}$OH) to more complex prebiotic compounds like Propyl Cyanide (C$_3$H$_7$CN) \citep{bell09} and ethanolamine (NH$_2$CH$_2$CH$_2$OH) \citep{Rivilla21}. G10.47+0.03 (hereafter, G10) is one of the notable hot cores located at a distance of ~10.7 kpc from Earth \citep{Urquhart2018}. With a bolometric luminosity of approximately $7\times10^{5}~L_\odot$ \citep{cesa10}, G10 ranks among the most luminous hot cores in the Galaxy. This source has been observed with various ground-based facilities, including ALMA, SMA, and the VLA, and the results have been reported \citep{rolf11,Gorai2020,Mondal21,mondal2023}.

The Herschel satellite was the first telescope designed for systematic investigation of the 500-5000 GHz (60-600 $\mu$m) frequency range \citep{Pilbratt10}, while the Heterodyne Instrument for the Far Infrared (hereafter HIFI) was the first high-resolution spectrometer that allowed a comprehensive study of the 480-1907 GHz spectrum using heterodyne methods \citep{Graauw10}. The Herschel Key Project on EXtra-Ordinary Sources (HEXOS; \citep{Bergin10}) was developed to investigate the chemical composition of massive star-forming regions \citep{Ceccarelli2010, Luca12,van13}. In this project, \cite{Crockett14}, \cite{Tahani16}, and \cite{Nagy17} conducted spectral line surveys at high frequencies, which are challenging to obtain through ground-based observations. They focused on Orion-KL, Orion South (referred to as Orion-S), and the Orion Bar, all of which are located in the Orion A Molecular Cloud, approximately 420 parsecs away\citep{Menten07}.
The HEXOS spectral surveys of Orion KL and Sgr B2 provide essential benchmarks for line identification, excitation analysis, and chemical complexity in massive star-forming regions. The line assignments and excitation conditions derived for G10 are therefore crucial for placing this source in context with these well-studied regions.
However, Sgr B2 lies in the Central Molecular Zone, whereas G10 is a Galactic-disk hot molecular core with distinct physical conditions, kinematics, and evolutionary state. G10 hosts ultracompact H II regions and deeply embedded massive protostars, placing it at a different evolutionary stage from Orion KL. Therefore, G10 cannot be treated as a simple scaled analogue of either CMZ sources such as Sgr B2 or nearby protoclusters like Orion KL.
Earlier submillimeter observations \citep{mondal2023,Mondal21,gora20} revealed several COMs in G10. A high-frequency survey of G10 is thus essential to characterize the chemical and physical properties of this source across the submillimeter to far-infrared regime.

In this study, we present an in-depth analysis of the HIFI spectrum survey, focusing on G10 at high frequencies that are typically inaccessible to ground-based telescopes due to atmospheric absorption. The instrument detected a wide range of molecular lines, enabling a more in-depth understanding of G10's chemical complexity and physical conditions. This large data set is crucial for expanding our knowledge of star formation and the role of complex compounds in these processes.

\section{Methodology \label{sec:method}}
\subsection{Observations}
G10 has been observed with the HIFI, one of the three instruments of the Herschel Space Observatory \citep{Pilbratt10}. The data reported in this work were taken from the Herschel Science Archive\footnote{\url{https://archives.esac.esa.int/hsa/whsa/}} (ID: OT\_1pschilke\_2, PI: P. Schilke, Dated: 12-10-2012).  A summary of the observations is given in Table \ref{tab: Observation Details}. The observations were performed covering five spectral ranges: (i) 514.7735-547.0775 GHz, (ii) 969.884-1001.195 GHz, (iii) 1020.363-1039.89 GHz, (iv) 1146.837-1181.16 GHz, and (v) 1205.8395-1226.5315 GHz. Each receiver band has separate channels for horizontal and vertical polarizations, equipped with a dedicated wide-band spectrometer with a native spectral resolution of 1.1 MHz \citep{Roelfsema12}. The systematic velocity of this source has taken as 67 km s$^{-1}$ \citep{rolf11, Gorai2020, mondal2023}. 
All observations were conducted in Dual Beam Switch mode using the Fast Chop option, with a chopping frequency greater than 0.5 Hz. The noise of 1$\sigma$ rms (0.035~K) is determined from the line-free regions of the spectrum adjacent to the lines.
We obtained different RA and Dec for each observation, which are slightly shifted from the others. Across 515-1230 GHz, the expected FWHM is 40 -18$^{\prime\prime}$, whereas the pointing errors
of $\leq$ 1–2$^{\prime\prime}$ are therefore a small fraction (~2–5\% or less) of the beam.

\begin{table}[h]
\centering
\tiny
\caption{Details of the HERSCHEL/HIFI spectral survey towards the region G10.\label{tab: Observation Details}}
\vskip 0.2cm
\begin{tabular}{ccccc}\\
\hline
 Observation Id &  RA (in J2000) &  Dec (in J2000) &  Band &  Frequency Range (GHz)\\
 \hline
1342242817&18$^{h}$08$^{m}$38.19$^{s}$&-19$^      0$51$^{'}$49.55$^{''}$& 1a&514.7735-	
547.0775 \\
\hline
1342244050& 	
18$^{h}$08$^{m}$38.22$^{s}$&	
-19$^{0}$51$^{'}$50.23$^{''}$ & 	
4a&969.884-1001.195 \\
\hline
1342244051&	
18$^{h}$08$^{m}$38.22$^{s}$ &	
-19$^{0}$51$^{'}$50.18$^{''}$ & 	
4a&1020.363-	
1039.89 	
\\
\hline
1342253155&	
18$^{h}$08$^{m}$38.27$^{s}$& 	
-19$^{0}$51$^{'}$49.32$^{''}$&5a &	
1146.837-	
1181.16 \\
\hline	
1342253156& 	
18$^{h}$08$^{m}$38.26$^{s}$&	
-19$^{0}$51$^{'}$49.41$^{''}$ &	
5a & 1205.8395-1226.5315\\
\hline
\end{tabular}
\end{table}

\subsection{Line identification}
Line identification is carried out using the CASSIS spectrum analyzer \footnote{\url{http://cassis.irap.omp.eu}  \citep{vast15}}. We consider unblended lines with a signal-to-noise ratio greater than the 3$\sigma$ limit. Line parameters were obtained using a single Gaussian ﬁt to the observed spectral proﬁle of each unblended transition. The process involved using the Jet Propulsion Laboratory \citep[JPL,][]{pick98}\footnote{\url{(http://spec.jpl.nasa.gov/)}} and the Cologne Database for Molecular Spectroscopy \citep[CDMS,][]{mull01,mull05}\footnote{\url{(https://www.astro.uni-koeln.de/cdms)}} spectral line databases to identify molecular species in our observed spectra. First, we identified the lines by visually evaluating each spectral characteristic in the HIFI bands and compared their transition frequencies to those specified in databases. We examined line blending, systematic velocity (VLSR), upper state energy (E$_u$), and the Einstein coefficient (A$_{ij}$) to confirm the identification of a molecular transition that corresponded to the spectra. Finally, we used local thermodynamic equilibrium (LTE) modeling to confirm or reject the assigned molecular species for the observed spectral feature.

We further evaluated all these transitions with non-LTE RADEX calculations \citep{vandertak2007} to determine their optical depth. For transitions found to be optically thin, we then employ Markov Chain Monte Carlo (MCMC) techniques within a local thermodynamic equilibrium (LTE) framework to generate synthetic spectra, enabling a quantitative assessment and validation of our molecular line identifications.

\subsection{SED fitting}
To complement the molecular line analysis, we performed a spectral energy distribution (SED) fitting for G10. This approach provides estimates of key physical parameters such as dust temperature, luminosity, and envelope mass, which cannot be directly obtained from line spectroscopy alone. These global properties serve as critical boundary conditions for interpreting the chemical complexity revealed by the HIFI survey.
The spectral energy distribution (SED) analysis for our sample was carried out with the latest release (v0.9.11) of \texttt{sedcreator}, an open-source Python package detailed in \cite{Fedriani2023} and further discussed in \cite{Telkamp2025}. The package is available on both GitHub\footnote{\url{https://github.com/fedriani/sedcreator}} and PyPi\footnote{\url{https://pypi.org/project/sedcreator/}}, with full documentation at \url{https://sedcreator.readthedocs.io/}. \texttt{sedcreator} provides two primary classes: \texttt{SedFluxer}, which performs aperture photometry on specified images, coordinates, and aperture sizes using functions from \cite{bradley2020}; and \texttt{SedFitter}, which fits observations to a grid of models based on the radiative transfer framework of \cite{Zhang2018}. 

\section{Results and Discussions\label{sec:result}}
\subsection{Physical properties of G10 \label{sec:sed}}
\begin{figure}[h]
   \centering 
\includegraphics[scale=1]{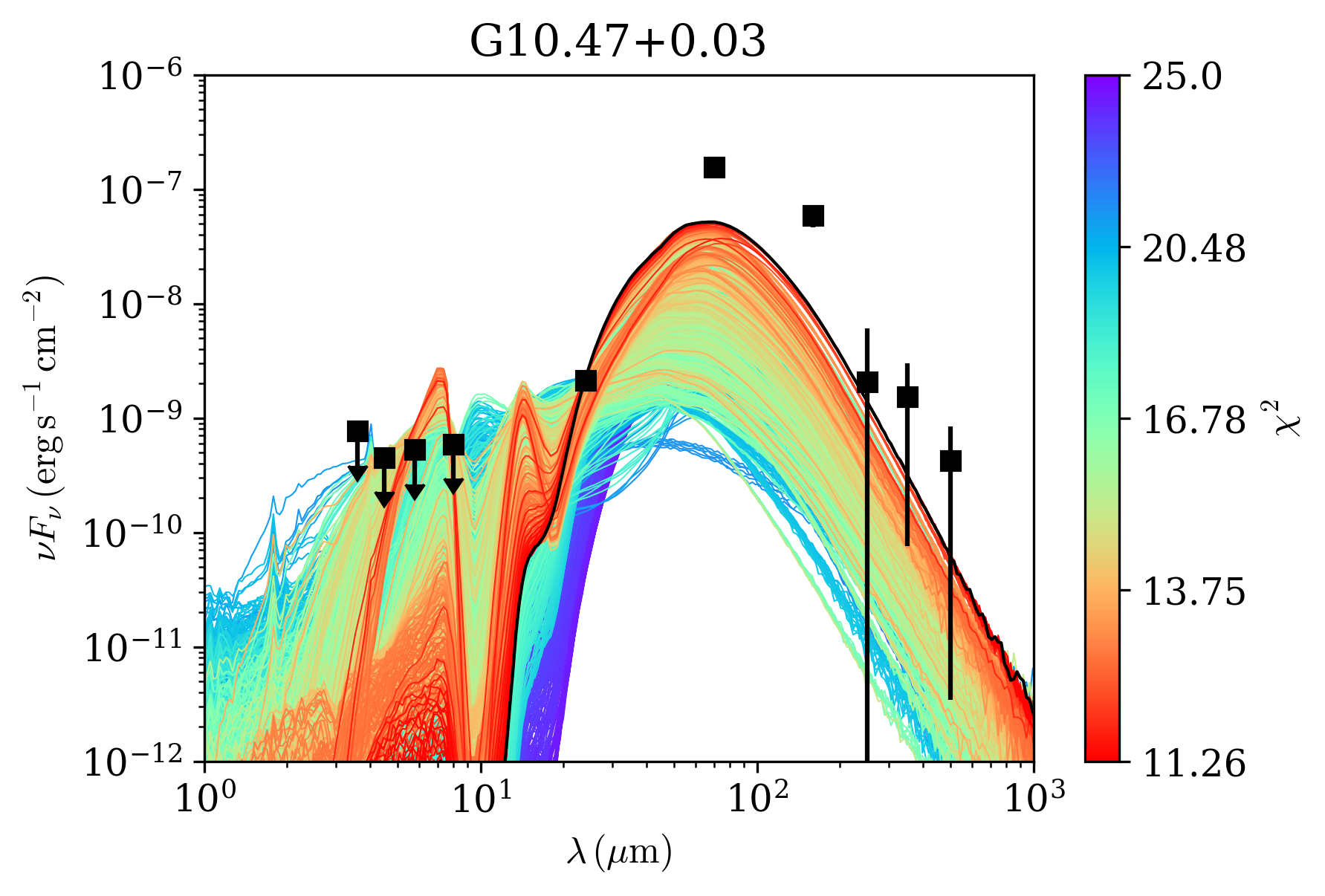}
    \caption{SED fitting of G10. The fixed-aperture, background-subtracted SED data were fitted using the \citep{Zhang2018} protostellar model grid, and the best-fitting model is shown as a black line while all other``good” model fits (see text) are shown with colored lines (red to blue with increasing $\chi^2$). We define good models as those with $\chi^2$ values up to twice the minimum $\chi^2$ value, i.e., $\chi^2 < 2\,\chi^2_{\min}$ \citep{Fedriani2023,Telkamp2025}.
}
    \label{fig:sed}
\end{figure}

\begin{figure}[h]
    \includegraphics[width=\linewidth]{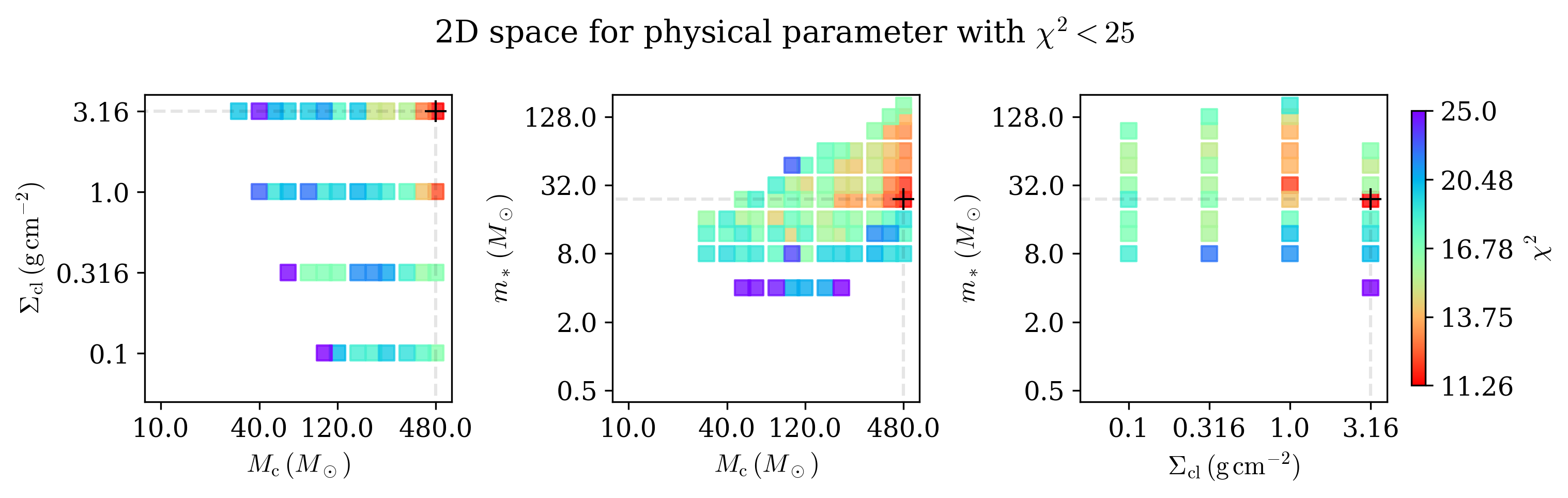}
    \caption{Comparison of key model parameters for each source: 
$\Sigma_{\mathrm{cl}}$ vs.\ $M_{\mathrm{c}}$ (left column), 
$m_{*}$ vs.\ $M_{\mathrm{c}}$ (center column), and 
$m_{*}$ vs.\ $\Sigma_{\mathrm{cl}}$ (right column). 
Only the ``good'' model fits are shown, color-coded by their $\chi^{2}$ values. 
The black cross marks the best-fitting model.}
    \label{fig:sed-para}
\end{figure}

The physical properties of the source G10 have been investigated in various works. The column density of the source, based on recent ALMA observations, has been reported to be $\rm 1.35\times10^{25}\ cm^{-2}$, and the excitation temperature is around 150--400 K \citep{Gorai2020, Mondal21,mondal2023}. \cite{van13} reported a modeled envelope mass of 1168~M$_\odot$ and a luminosity of $7\times10^{5}$~L$_\odot$, based on an envelope size of 60000~AU and a source distance of 5.8~kpc.
For SED fitting, we used archival data from Spitzer/IRAC \citep{Werner2004, Fazio2004} at 3.6, 4.5, 5.8, and 8.0 $\mu$m from the Spitzer Heritage Archive, as well as Herschel/PACS and SPIRE \cite{Griffin2010} at 70, 160, 250, 350, and 500 $\mu$m  from the ESA Herschel Science Archive. We also included MIPS 24 $\mu$m data. Herschel 70 $\mu$m data were used to automatically determine the optimal aperture using specialised functions from sedcreator \citep{Fedriani2023}, which was found to be 18 arcsec. This aperture was then applied as a fixed aperture to derive the fluxes from images at different wavelengths. In addition, we used an annulus of inner radius equal to the aperture radius and outer radius equal to 2 $\times$ aperture radius to subtract the background from the fluxes. Figure \ref{fig:sed} shows the SED fitting, while Figure \ref{fig:sed-para} displays the 2D model fitting parameters of the source. The best-fit SED results are summarized in Table \ref{tab:sed_params}. Model parameters and derived quantities such as $\chi^2$ is the chi-squared goodness-of-fit statistic, $M_c$ is the core mass, $\Sigma_{\rm cl}$ is the mass surface density of the surrounding clump, $R_c$ is the core radius, $m_\ast$ is the current stellar mass, $\theta_{\rm view}$ is the viewing angle measured from the outflow axis, $A_V$ is the visual extinction along the line of sight, $M_{\rm env}$ is the envelope mass, $\theta_{\rm w,esc}$ is the half-opening angle of the outflow cavity (escape angle),  $\dot{M}_{\rm disk}$ is the disk mass accretion rate onto the star, $L_{\rm bol,iso}$ is the isotropic bolometric luminosity, and $L_{\rm bol}$ is the true bolometric luminosity of the source. Our SED-derived luminosity is comparable to the previously reported value \citep{vanderTak2013}. 
From the mass surface density ($\rm{\Sigma}_{cl}$ ), we can estimate the molecular hydrogen column density (N$_{\mathrm H_2}$) using the following relation:
\[
N_{\mathrm{H_2}} = \frac{\Sigma_{\rm cl}}{\mu \, m_{\rm H}}.
\]
Here, $m_{\rm H}$ is the hydrogen atomic mass and $\mu = 2.8$ is the mean molecular weight per H$_2$ molecule (including helium). The value of $\rm{\Sigma}_{cl}$ obtained from our best fitted model is $3.16$, yields N$_{\mathrm{H_2}}= 6.74\times10^{23}$ cm$^{-2}$. This value is slightly lower than that reported in previous ALMA observations. This difference arises from the larger spatial coverage achieved using the optimal aperture method, which encompasses an extended envelope in our analysis. This extended envelope is also probed by Herschel observations, whereas earlier column density measurements were based on high-resolution ALMA observations and focused on the compact, dense core.

\begin{table}[h!]
\tiny
\centering
\caption{Best fitted SED parameters.}
\label{tab:sed_params}
\vskip 0.5cm
\begin{tabular}{c c c c c c c c c c c c }
\hline
$\chi^2$ &
$M_c$ &
$\Sigma_{\mathrm{cl}}$ &
$R_c$ &
$m_\ast$ &
$\theta_{\mathrm{view}}$ &
$A_V$ &
$M_{\mathrm{env}}$ &
$\theta_{\mathrm{w,esc}}$ &
$\dot{M}_{\mathrm{disk}}$ &
$L_{\mathrm{bol,iso}}$ &
$L_{\mathrm{bol}}$ \\
&
($M_\odot$) &
(g\,cm$^{-2}$) &
(pc) &
($M_\odot$) &
(deg) &
(mag) &
($M_\odot$) &
(deg) &
($M_\odot\,\mathrm{yr}^{-1}$) &
($L_\odot$) &
($L_\odot$) \\
\hline
6.76 & 480.00 & 3.16 & 0.09 & 24.00 & 82.82 & 56.13 & 440.54 & 11.50 & 1.95e-03 & 2.43e+05 & 2.94e+05 \\
8.21 & 480.00 & 1.00 & 0.16 & 32.00 & 22.33 & 160.02 & 414.30 & 18.51 & 9.31e-04 & 3.72e+05 & 2.95e+05 \\
7.23 & 400.00 & 3.16 & 0.08 & 24.00 & 85.70 & 67.09 & 361.65 & 12.84 & 1.86e-03 & 2.34e+05 & 3.01e+05 \\
10.08 & 400.00 & 1.00 & 0.15 & 32.00 & 22.33 & 218.17 & 329.80 & 20.63 & 8.74e-04 & 7.17e+05 & 2.85e+05 \\
10.09 & 480.00 & 1.00 & 0.16 & 64.00 & 12.84 & 333.10 & 324.63 & 31.74 & 1.21e-03 & 4.75e+06 & 8.41e+05 \\
\hline
\end{tabular}\\
Note- The best five models taken from the 432 physical models for the non-restricted
$\rm{\Sigma}_{cl}$ case. $\chi^2$ is the chi-squared goodness-of-fit statistic, $M_c$ is the core mass, $\Sigma_{\rm cl}$ is the mass surface density of the surrounding clump, $R_c$ is the core radius, $m_\ast$ is the current stellar mass, $\theta_{\rm view}$ is the viewing angle measured from the outflow axis, $A_V$ is the visual extinction along the line of sight, $M_{\rm env}$ is the envelope mass, $\theta_{\rm w,esc}$ is the half-opening angle of the outflow cavity (escape angle),  $\dot{M}_{\rm disk}$ is the disk mass accretion rate onto the star, $L_{\rm bol,iso}$ is the isotropic bolometric luminosity, and $L_{\rm bol}$ is the true bolometric luminosity of the source.
\end{table}

\subsection{Identified species}
\begin{figure}[h]
\centering
\includegraphics[width=\linewidth]{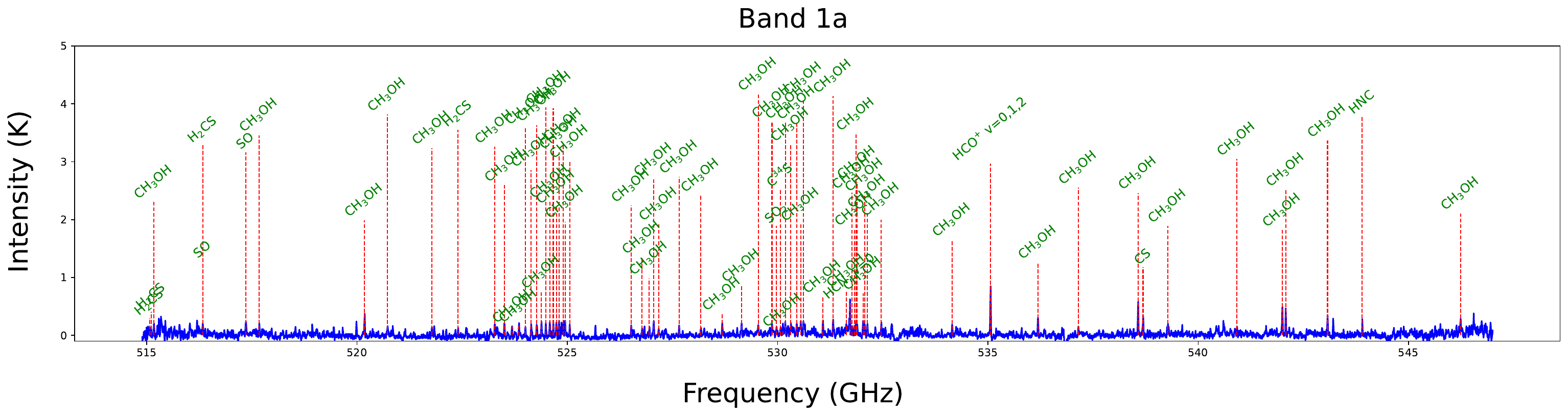}\\
\includegraphics[width=\linewidth]{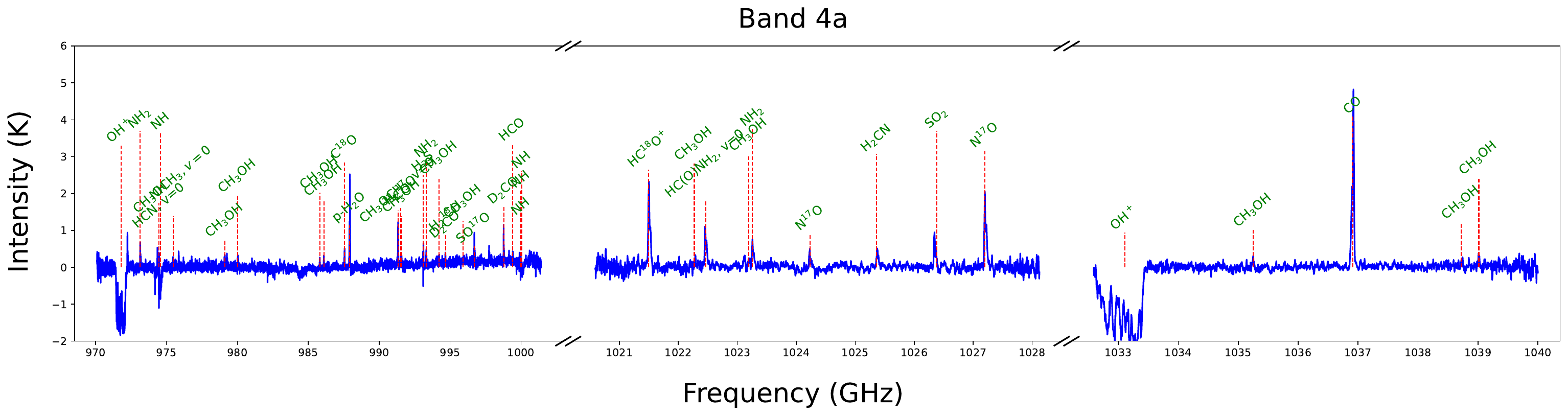}\\
\includegraphics[width=\linewidth]{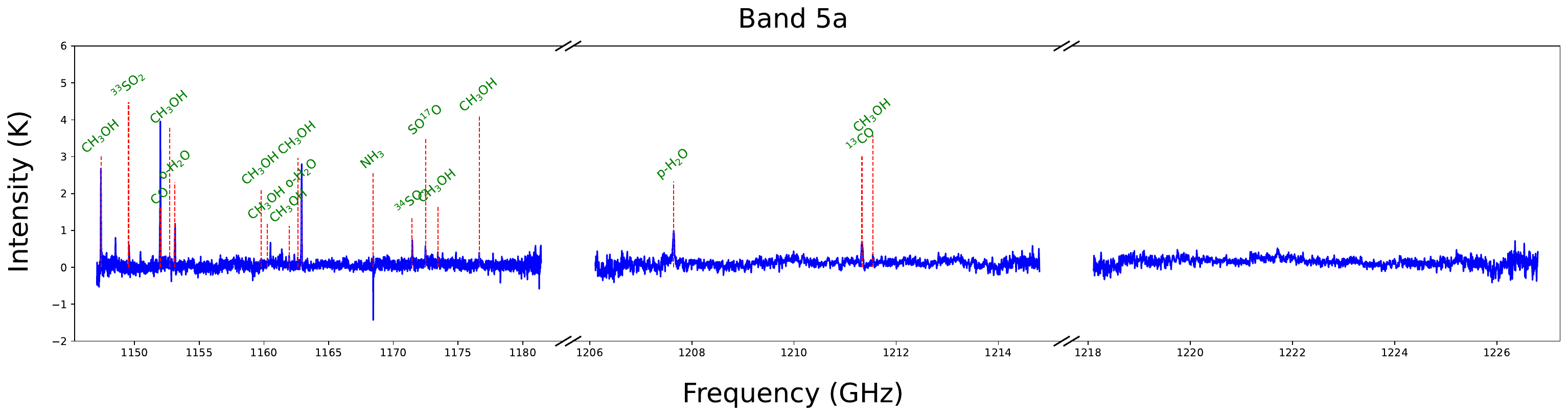}
\caption{G10 Spectra with identified species.}
\label{fig: spectra-G10}
\end{figure}

Here, we present the results of the molecular line survey conducted with HIFI toward the hot molecular core G10. The broad spectral coverage of HIFI enables a comprehensive investigation of molecular emission, ranging from simple diatomic species to complex organic molecules, thereby highlighting the remarkable chemical diversity of this source. The species identified in this survey are listed in Table \ref{tab: Species} and shown in Figure \ref{fig: spectra-G10}, which reflects the rich inventory characteristic of hot cores.
\begin{table}[h!]
\tiny
\centering
\caption{List of species identified in G10 from this survey. \label{tab: Species}}
\vskip 0.5cm
\begin{tabular}{c c c}		
\hline								
Diatomic Molecules   &	Multi-atomic Molecules	& Ionized\\	\hline\hline \\	
CO&HCN, v=0&HCO$^+$, v=0,1,2\\
$^{13}$CO& H$_2$$^{18}$O&OH$^+$$^b$\\
C$^{18}$O&o${-}$H$_2$O&HC$^{18}$O$^{+}$$^a$\\
CS&p${-}$H$_2$O&...\\
C$^{34}$S&H$_{2}$CS&...\\
SO&CH$_{3}$OH, vt=0-2&...\\
N$^{17}$O$^a$&....&...\\
NH$^b$&NH$_2$&...\\
...&SO$_{2}$&...\\
...&HNC&...\\
...&CH$_3$OCH$_3$, v=0$^a$&...\\
...&H$_2$$^{17}$O&....\\
...&H$_2$S$^b$&.....\\
...&D$_2$CO$^a$&...\\
...&HC(O)NH$_2$, v=0$^a$&...\\
...&NH$_3$$^b$&...\\
...&$^{33}$SO$_2$$^a$&...\\
...&$^{34}$SO$_2$$^a$&...\\
...&SO$^{17}$O$^a$&...\\
...&HCO$^a$&...\\

\hline	
 \end{tabular}	

         $^a$ Below three sigma or blended with other transitions,
         $^b$ Observed absorption profiles.  
      
  \end{table}
  
To provide quantitative constraints, all detected transitions are catalogued in Table \ref{tab:line_parameters}, including their rest frequencies, upper-state energies ($E_u$), line widths (FWHM), peak intensities, and integrated intensities ($\int T_{\rm mb} dv$), where ${T_{mb}}$ is the main beam temperature obtained from the observations in Kelvin and $v$ is the FWHM in km s$^{-1}$. This compilation serves as the basis for subsequent analyses, including rotational diagram modeling and LTE radiative transfer calculations, aimed at deriving column densities and excitation conditions for the identified species.

The observed molecules serve as key tracers of various physical properties of the present source. CO and its isotopologues trace large-scale gas and outflows, while species such as HCN, H$_2$O, HCO$^{+}$, CS, and SO probe dense and warm regions associated with infall and shock activity. The detection of complex organic molecules such as CH$_3$OH and H$_2$CS indicates the presence of high-temperature, high-density gas, typical of hot molecular cores.

\subsubsection{Column density estimation} 
{\hskip 6.5 cm {\it Rotational diagram analysis}}\\
This analysis assumes that the detected transitions are optically thin in LTE.
For optically thin lines, the column density can be calculated using the expression \citep{gold99} as\\
\begin{equation}
\frac{N_u^{thin}}{g_u}=\frac{3k_B\int{T_{mb}dv}}{8\pi^{3}\nu S\mu^{2}},
\end{equation}
where $g_u$ is the degeneracy of the upper state, $k_B$ is the Boltzmann constant, ${\int T_{mb}dv}$ is the integrated intensity,
$\nu$ is the rest frequency, $\mu$ is the electric dipole moment, and $S$ is the transition line strength.\\
The total column densities $N_{total}$ under LTE conditions can be written as,
\begin{equation}
\frac{N_u^{thin}}{g_u}=\frac{N_{total}}{Q(T_{rot})}e^{(-E_u/k_BT_{rot})},
\label{eqn:Ntotal}
\end{equation}
where $E_u$ is the upper-state energy and ${Q(T_{rot})}$ is the partition function at rotational temperature $T_{rot}$.\\
Using the logarithm of both sides of the above equation \ref{eqn:Ntotal}, it gives 
\begin{equation}
log\Bigg(\frac{N_u^{thin}}{g_u}\Bigg)=-\Bigg(\frac{log\ e}{T_{rot}}\Bigg)\Bigg(\frac{E_u}{k_B}\Bigg)+log\Bigg(\frac{N_{total}}{Q(T_{rot})}\Bigg).
\label{eqn:rd}
\end{equation}
This expresses a linear relationship in logarithmic space between the upper-state energy and the column density of the corresponding level, allowing both the excitation temperature and total column density of a species to be derived simultaneously through rotational diagram analysis.

\begin{figure}[h]
    \centering
    \includegraphics[scale=0.5]{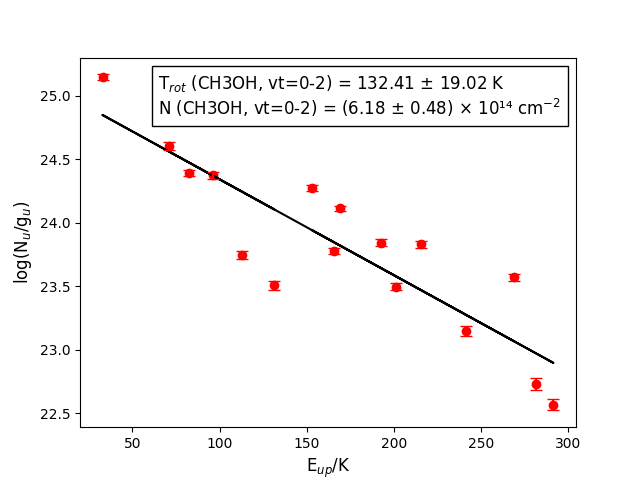}
    \includegraphics[scale=0.5]{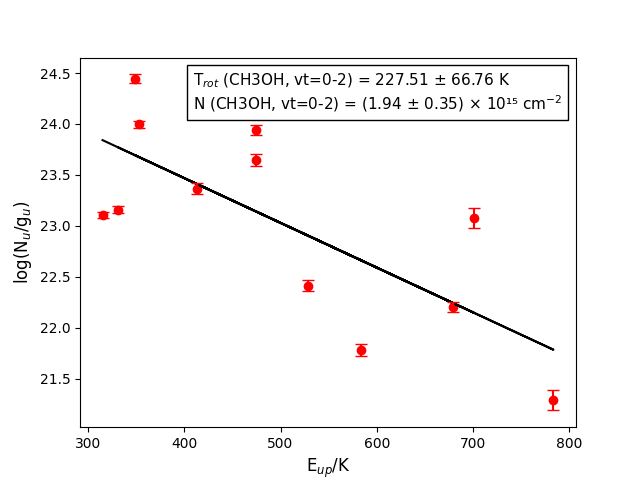}
    \caption{
    Above two diagrams present the rotational spectrum for transitions occurring below and above 300 K for CH$_3$OH (vt=0-2). The vertical bars indicate the error margins for the measurements in each panel. Each panel provides the best-fitted values for the rotational temperature and column density.}
    \label{fig: RD_E-CH3OH}
\end{figure}

Among the detected species, methanol (CH$_3$OH, vt=0-2) and thioformaldehyde (H$_2$CS) exhibit multiple transitions. However, in the case of H$_2$CS, four transitions were identified, two of which show significantly narrower line widths compared to the others, making a consistent rotational analysis unreliable. We therefore restrict our rotational diagram analysis to methanol, from which excitation temperatures and column densities were derived (see Figure \ref{fig: RD_E-CH3OH}).

\tiny
\setlength{\LTcapwidth}{\textwidth}
\begin{longtable}{|p{1.6cm}|p{4cm}|p{2cm}|p{1.2cm}|p{2cm}|p{2cm}|p{2cm}|}
\caption{Summary of the Line Parameters of Observed Molecules toward G10.
\label{tab:line_parameters}}\\
\hline
Species & Quantum number & Frequency (GHz) & $E_u$ (K) &
FWHM (km~s$^{-1}$) & Peak intensity (K) &
$\int T_{mb}\,dv$ (K~km~s$^{-1}$) \\
\hline
\endfirsthead
\multicolumn{7}{c}{\textit{Table \thetable\ continued from previous page}} \\
\hline
Species & Quantum number & Frequency (GHz) & $E_u$ (K) &
FWHM (km~s$^{-1}$) & Peak intensity (K) &
$\int T_{mb} dv$ (K~km~s$^{-1}$) \\
\hline
\endhead

\hline
\multicolumn{7}{r}{\textit{Continued on next page}} \\
\endfoot

\hline
\endlastfoot
& $(2_{2,0,2}-1_{1,1,2})$     & 520.179054$^{*}$ & 32.86  & 7.69$\pm$ 0.29 & 0.378 $\pm$ 0.006 & 3.099 $\pm$ 0.173 \\
CH$_3$OH,& $(12_{4,9,2}-12_{3,9,2})$ & 524.384667 & 268.91 & 9.08 $\pm$ 0.23 & 0.234 $\pm$ 0.003 & 2.264 $\pm$ 0.088 \\
vt=0-2& $(11_{4,8,2}-11_{3,8,2})$ & 524.488918 & 241.07 & 5.76 $\pm$ 0.16 & 0.220 $\pm$ 0.003 & 1.348 $\pm$ 0.054 \\
& $(10_{4,7,2}-10_{3,7,2})$ & 524.582604 & 215.55 & 9.55 $\pm$ 0.34 & 0.235 $\pm$ 0.003 & 2.388 $\pm$ 0.123 \\
& $(9_{4,6,2}-9_{3,6,2})$   & 524.666219 & 192.34 & 8.63 $\pm$ 0.40 & 0.232 $\pm$ 0.004 & 2.134 $\pm$ 0.135 \\
& $(7_{4,4,2}-7_{3,4,2})$   & 524.804977 & 152.89 & 8.57 $\pm$ 0.41 & 0.258 $\pm$ 0.003 & 2.353 $\pm$ 0.148 \\
& $(14_{2,13,0}-14_{1,14,0})$ & 526.520131 & 281.29 & 5.80 $\pm$ 0.31 & 0.179 $\pm$ 0.004 & 1.108 $\pm$ 0.084 \\
& $(14_{3,12,1}-14_{2,13,1})$ & 527.658026 & 291.46 & 6.52 $\pm$ 0.51 & 0.173 $\pm$ 0.005 & 1.202 $\pm$ 0.133 \\
& $(11_{3,9,1}-11_{2,10,1})$  & 529.143188 & 200.92 & 10.48 $\pm$ 0.53 & 0.210 $\pm$ 0.004 & 2.339 $\pm$ 0.164 \\
& $(11_{1,10,1}-10_{1,9,1})$  & 532.031395$^{*}$& 174.26 & 9.15 $\pm$ 0.23 & 0.259 $\pm$ 0.003 & 2.525 $\pm$ 0.093 \\
& $(11_{2,10,1}-10_{2,9,1})$  & 532.466250$^{*}$ & 175.53 & 8.29 $\pm$ 0.17 & 0.235 $\pm$ 0.002 & 2.070 $\pm$ 0.062 \\
& $(11_{1,10,0}-10_{1,9,0})$  & 536.191058$^{*}$ & 169.01 & 10.31 $\pm$ 0.37 & 0.307 $\pm$ 0.004 & 3.365 $\pm$ 0.165 \\
& $(10_{2,8,0}-9_{1,9,0})$    & 986.097819 & 165.40 & 8.79 $\pm$ 0.65 & 0.300 $\pm$ 0.013 & 2.810 $\pm$ 0.341 \\
& $(8_{3,6,1}-8_{2,7,1})$     & 530.123296 & 131.28 & 7.49$\pm$ 0.36 & 0.206 $\pm$ 0.004 & 1.645 $\pm$ 0.110 \\
& $(7_{3,5,1}-7_{2,6,1})$     & 530.316194 & 112.71 & 7.71$\pm$ 0.32 & 0.218 $\pm$ 0.002 & 1.786 $\pm$ 0.092 \\
& $(6_{3,4,1}-6_{2,5,1})$     & 530.454670 & 96.46  & 12.72 $\pm$ 0.31 & 0.205 $\pm$ 0.002 & 2.769 $\pm$ 0.094 \\
& $(5_{3,3,1}-5_{2,4,1})$     & 530.549228 & 82.53  & 8.27 $\pm$ 0.17 & 0.252 $\pm$ 0.002 & 2.220 $\pm$ 0.062 \\
& $(4_{3,2,1}-4_{2,3,1})$     & 530.610267 & 70.93  & 7.31 $\pm$ 0.28 & 0.251 $\pm$ 0.003 & 1.956 $\pm$ 0.102 \\
& $(5_{1,5,0}-4_{0,4,0})$     & 538.570553$^{*}$ & 49.06  & 8.43 $\pm$ 0.23& 0.582 $\pm$ 0.005 & 5.218 $\pm$ 0.190 \\
& $(6_{3,4,0}-5_{2,3,0})$     & 542.000954$^{*}$ & 98.55  & 9.37$\pm$ 0.14 & 0.495 $\pm$ 0.002 & 4.938 $\pm$ 0.104 \\
& $(6_{3,3,0}-5_{2,4,0})$     & 542.081952$^{*}$ & 98.55  & 7.21 $\pm$ 0.18 & 0.473 $\pm$ 0.004 & 3.634 $\pm$ 0.130 \\
& $(8_{0,8,1}-7_{1,7,2})$     & 543.076175$^{*}$ & 96.61  & 9.24 $\pm$ 0.31 & 0.351 $\pm$ 0.004 & 3.455 $\pm$ 0.157 \\
& $(16_{0,16,0}-15_{1,15,0})$ & 515.170245 & 315.21 & 10.21 $\pm$ 0.29 & 0.209 $\pm$ 0.002 & 2.273 $\pm$ 0.093 \\
& $(19_{7,12,2}-20_{6,14,2})$ & 516.922004 & 700.76 & 4.99 $\pm$ 0.39& 0.103 $\pm$ 0.005 & 0.546 $\pm$ 0.073 \\
& $(16_{2,15,0}-15_{3,12,0})$ & 517.676592 & 353.12 & 7.82 $\pm$ 0.28 & 0.213 $\pm$ 0.004 & 1.770 $\pm$ 0.105 \\
& $(13_{5,9,0}-14_{4,10,0})$ & 521.776019 & 349.05 & 10.50 $\pm$ 0.66& 0.142 $\pm$ 0.003 & 1.592 $\pm$ 0.142 \\
& $(24_{4,21,2}-24_{3,21,2})$ & 522.046360 & 783.40 & 4.68 $\pm$ 0.32 & 0.094 $\pm$ 0.003 & 0.467 $\pm$ 0.048 \\
& $(19_{4,16,2}-19_{3,16,2})$ & 523.306240 & 528.59 & 8.05 $\pm$ 0.28 & 0.134 $\pm$ 0.002 & 1.144 $\pm$ 0.056 \\
& $(21_{3,19,1}-21_{2,20,1})$ & 529.038126 & 583.96 & 5.59 $\pm$ 0.41 & 0.136 $\pm$ 0.003 & 0.809 $\pm$ 0.079 \\
& $(11_{8,4,2}-10_{8,3,2})$  & 531.437209 & 474.56 & 7.23 $\pm$ 0.25 & 0.129 $\pm$ 0.002 & 0.994 $\pm$ 0.053 \\
& $(11_{8,3,0}-10_{8,2,0})$  & 531.460073 & 474.06 & 9.15 $\pm$ 0.35 & 0.137 $\pm$ 0.002 & 1.337 $\pm$ 0.072 \\
& $(6_{2,5,5}-7_{3,5,5})$    & 537.151571 & 413.23 & 6.55 $\pm$ 0.24 & 0.153 $\pm$ 0.003 & 1.068 $\pm$ 0.062 \\
& $(17_{1,17,3}-17_{2,15,3})$ & 545.755816 & 678.86 & 3.55 $\pm$ 0.09& 0.217 $\pm$ 0.002 & 0.820 $\pm$ 0.028 \\
& $(14_{4,11,2}-14_{3,11,2})$ & 524.142239 & 331.53 & 8.95 $\pm$ 0.90 & 0.186 $\pm$ 0.005 & 1.771 $\pm$ 0.232 \\
\hline
CO & (10 - 9) & 1151.985452 & 304.16 & $7.98 \pm 0.27$ & $3.731 \pm 0.085$ & $31.68 \pm 1.30$ \\
~ & (9 - 8) & 1036.912393 & 248.88 & $6.31 \pm 0.10$ & $4.803 \pm 0.054$ & $32.27 \pm 0.63$ \\ \hline
$^{13}$CO  & (11 - 10) & 1211.329661 & 348.92 & $9.31 \pm 0.22$ & $0.662 \pm 0.011$ & $6.56 \pm 0.20$ \\
~ & (9 - 8) & 991.329305 & 237.93 & $8.74 \pm 0.07$ & $1.246 \pm 0.008$ & $11.59 \pm 0.12$ \\ \hline
C$^{18}$O  & \tiny{\hspace{0.05em} $(9-8)$} & 987.56040&237.03&$8.27\pm0.26$& $0.485\pm0.013$& $4.27\pm0.19$ \\ \hline
CS & (11 - 10) & 538.68883 & 155.15 & $11.70 \pm 0.40$ & $0.300 \pm 0.006$ & $3.73 \pm 0.15$ \\ \hline
C$^{34}$S & (11 - 10) & 530.071221 & 152.66 & $9.12 \pm 1.73$ & $0.158 \pm 0.005$ & $1.54 \pm 0.30$ \\ \hline
SO & $(12_{1,3}-11_{1,2})$ & 517.3543051 & 165.78 & $14.07 \pm 0.32$ & $0.236 \pm 0.003$ & $3.53 \pm 0.11$ \\
~ & $(12_{1,2} - 11_{1,1})$ & 516.3354166 & 174.22 & $14.27 \pm 0.39$ & $0.185 \pm 0.004$ & $2.82 \pm 0.09$ \\ \hline
SO$_2$ & $(9_{4,6}-8_{3,5})$ & 529.97493 & 80.64 & $8.58 \pm 0.40$ & $0.147 \pm 0.004$ & $1.34 \pm 0.07$ \\ \hline
$^{33}$SO$_2$ & $(23_{4,20,24.5}-22_{1,21,23.5})$ & 1149.533678 & 292.80 & $10.13 \pm 0.49$ & $0.460 \pm 0.019$ & $4.95 \pm 0.32$ \\ \hline
$^{34}$SO$_2$ & $(26_{2,24}-25_{1,25})$ & 1171.444506 & 339.63 & $10.78 \pm 0.24$ & $0.701 \pm 0.012$ & $8.04 \pm 0.22$ \\ \hline
SO$^{17}$O & $(29_{5,24}-28_{4,25})$ & 996.700636 & 450.31 & $8.28 \pm 0.24$ & $0.829 \pm 0.018$ & $7.30 \pm 0.27$ \\
~ & $(25_{3,23}-24_{0,24})$ & 1172.520701 & 312.02 & $14.52 \pm 0.41$ & $0.533 \pm 0.009$ & $8.24 \pm 0.27$ \\ \hline
 o-H$_2$O & $(3_{1,2} - 2_{2,1})$ & 1153.126822 & 215.20 & $11.82 \pm 0.36$ & $1.099 \pm 0.025$ & $13.83 \pm 0.53$ \\
 ~ & $(3_{2,1} - 3_{1,2})$ & 1162.911593 & 271.01 & $10.26 \pm 0.26$ & $2.572 \pm 0.050$ & $28.09 \pm 0.89$ \\ \hline
p-H$_2$O & $(2_{0,2} - 1_{1,1})$ & 987.926764 & 100.85 & $12.29 \pm 0.44$ & $1.950 \pm 0.057$ & $25.52 \pm 1.18$ \\
~ & $(4_{2,2} - 4_{1,3})$ & 1207.638714 & 454.34 & $9.76 \pm 0.33$ & $0.941 \pm 0.018$ & $9.77 \pm 0.38$ \\ \hline
H$_{2}^{18}$O & \tiny{\hspace{0.05em} $2_{0,2}-1_{1,1}$} &994.675129&100.61&$11.75\pm0.37$& $0.385\pm0.058$& $4.82\pm0.74$ \\ \hline
H$_2$$^{17}$O & $(2_{0,2} - 1_{1,1})$ & 991.52013 & 100.72 & $10.07 \pm 0.51$ & $0.269 \pm 0.007$ & $2.88 \pm 0.16$ \\ \hline
HCN, v=0 & (6 - 5) & 531.7163479 & 89.32 & $10.65 \pm 0.84$ & $0.526 \pm 0.025$ & $5.97 \pm 0.55$ \\
~ & (11 - 10) & 974.4871998 & 280.67 & $2.88 \pm 0.14$ & $0.400 \pm 0.017$ & $1.23 \pm 0.08$ \\ \hline
HNC & (6 - 5) & 543.897386 & 91.37 & $5.42 \pm 0.10$ & $0.288 \pm 0.004$ & $1.66 \pm 0.04$ \\ \hline
OH$^{+}$ & $(1_{2,1.5} - 0_{1,0.5})$ & 971.8053 & 46.64 & ...... & ...... & ...... \\
~ & $(1_{1,0.5} - 0_{1,1.5})$ & 1033.1129 & 49.58 & ......& ...... & ......\\ \hline
NH & $(1_{2,2.5,2.5} - 0_{1,1.5,1.5})$ & 974.4708 & 46.77 & ...... & ...... & ......   \\ 
~ & $(1_{1,1.5,1.5} - 0_{1,1.5,2.5})$ & 1000.00102 & 47.99 & ...... & ...... &......\\
\hline
NH$_2$
~ & $(2_{2,0,0,7,2.5} - 2_{1,1,0,14,1.5})$ & 993.3229713 & 167.87 & $10.19 \pm 0.21$ & $0.473 \pm 0.006$ & $5.13 \pm 0.14$ \\
\hline
NH$_3$ & $(2_{1,0} - 1_{1,1})$ & 1168.452394 & 79.34 & ...... & ...... &...... \\ \hline
N$^{17}$O & $(10_{-1,10.5,10}- 9_{1,9.5,10})$ & 1024.234273 & 280.93 & $9.95 \pm 0.72$ & $0.374 \pm 0.021$ & $3.97 \pm 0.37$ \\
~ & $(10_{-1,10.5,12}-9_{-1,9.5,11})$ & 1027.197994 & 280.92 & $13.66 \pm 1.22$ & $1.526 \pm 0.078$ & $22.20 \pm 2.28$ \\ \hline
H$_2$S & $(3_{0,3} - 2_{1,2})$ & 993.108247 & 102.76 &......& ...... & ...... \\ \hline
H$_2$CS  & $(15_{3,13} - 14_{3,12})$ & 515.075721 & 316.22 & $4.32 \pm 0.11$ & $0.148 \pm 0.001$ & $0.68 \pm 0.05$ \\
~ & $(15_{3,12} - 14_{3,11})$ & 515.10836 & 316.22 & $7.92 \pm 0.99$ & $0.116 \pm 0.002$ & $0.98 \pm 0.14$ \\
~ & $(15_{2,13} - 14_{2,12})$ & 516.337197 & 250.71 & $13.43 \pm 1.51$ & $0.192 \pm 0.004$ & $2.74 \pm 0.32$ \\ 
~ & $(15_{1,14} - 14_{1,13})$ & 522.40314 & 213.87 & $9.93 \pm 0.48$ & $0.122 \pm 0.003$ & $1.29 \pm 0.07$ \\ \hline
HCO$^{+}$ v=0,1,2 & $(6_{0,0} - 5_{0,0})$ & 535.061402 & 89.88 & $9.36 \pm 0.20$ & $0.831 \pm 0.014$ & $8.27 \pm 0.22$ \\ \hline
HC$^{18}$O$^{+}$ & (12 - 11) & 1021.498294 & 318.72 & $7.40 \pm 0.11$ & $2.285 \pm 0.023$ & $17.99 \pm 0.32$ \\ \hline
HCO & $(10_{2,9,10.5,10} - 11_{1,10,11.5,11})$ & 999.402896 & 359.86 & $12.74 \pm 0.58$ & $0.409 \pm 0.008$ & $ 5.54 \pm 0.28$ \\ \hline
D$_2$CO & $(17_{3,15} - 16_{3,14})$ & 994.673348 & 479.38 & $13.38 \pm 0.28$ & $0.340 \pm 0.004$ & $4.84 \pm 0.13$ \\
~ & $(17_{6,11} - 16_{6,10})$ & 998.7898855 & 624.30 & $9.03 \pm 0.41$ & $0.872 \pm 0.026$ & $8.38 \pm 0.45$ \\ \hline
CH$_3$OCH$_3$, v=0 & $(16_{12,4,3} - 15_{11,4,3})$ & 975.469645 & 325.97 & $7.10 \pm 0.35$ & $0.243 \pm 0.008$ & $1.84 \pm 0.12$ \\
~ & $(23_{10,13,1} - 22_{9,13,1})$ & 991.343273 & 391.66 & $8.72 \pm 0.08$ & $1.246 \pm 0.008$ & $11.56 \pm 0.13$ \\ \hline
HC(O)NH$_2$, v=0 & $(22_{5,17} - 21_{4,18})$ & 1022.469579 & 332.60 & $14.58 \pm 1.19$ & $0.881 \pm 0.036$ & $13.68 \pm 1.25$\\\hline
\multicolumn{7}{l}{\footnotesize
$^{*}$ Optically thick transition.}
\end{longtable}

{\hskip 5cm {\it Markov Chain Monte Carlo (MCMC) method}}\\
To analyze the line profiles of the observed species toward the hot core G10, we employ the MCMC approach under the assumption of LTE. We have extracted the best-ﬁtted key physical parameters, including the column density, excitation temperature, line width (FWHM), optical depth, and systemic velocity (V$_{\rm LSR}$). 
We used the Python scripting interface of CASSIS to derive the best-fitting physical parameters of the source through a $\chi^{2}$ minimization between the modeled and observed spectra across $N$ transitions. The $\chi^{2}$ value was computed as  
\begin{equation}
\chi_{i}^{2} = \sum_{j=1}^{N_{i}} \frac{(I_{ij}^{\mathrm{obs}} - I_{ij}^{\mathrm{model}})^{2}}{(\mathrm{rms}_{i}^{2} + (I_{ij}^{\mathrm{obs}} \times \mathrm{cal}_{i})^{2})},
\end{equation}
where $I_{ij}^{\mathrm{obs}}$ and $I_{ij}^{\mathrm{model}}$ are the observed and modeled intensities for channel $j$ of transition $i$, $\mathrm{rms}_{i}$ is the spectral noise, and $\mathrm{cal}_{i}$ is the calibration uncertainty.  
The reduced $\chi^{2}$ was obtained as  
\begin{equation}
\chi_{\mathrm{red}}^{2} = \frac{1}{\sum_{i}^{N_{\mathrm{spec}}} N_{i}}{\sum_{i=1}^{N_{\mathrm{spec}}} \chi_{i}^{2}}.
\end{equation}

In the MCMC computation, initial parameter values were randomly selected within the user-defined range $[X_{\min}, X_{\max}]$. The step at iteration $l$ was defined as  
\begin{equation}
\theta_{l+1} = \theta_{l} + \alpha (v - 0.05),
\end{equation}
where $v$ is a random number between 0 and 1. The step amplitude $\alpha$ is expressed as $\alpha = \frac{k(X_{\max} - X_{\min})}{k'}$, with  
\[
k =
\begin{cases}
r_{c}, & l > c, \\
\dfrac{(r_c-1)}{c }l +1, & l < c,
\end{cases}
\]
where $c$ and $r_{c}$ are the cutoff and cutoff ratio, respectively, while $k'$ is a reduced physical parameter.  
This approach ensures larger steps at the beginning for efficient parameter exploration and smaller steps near convergence to accurately determine the optimal $\chi^{2}$.

\begin{figure}[h]
    \centering
	\includegraphics[width=\linewidth]{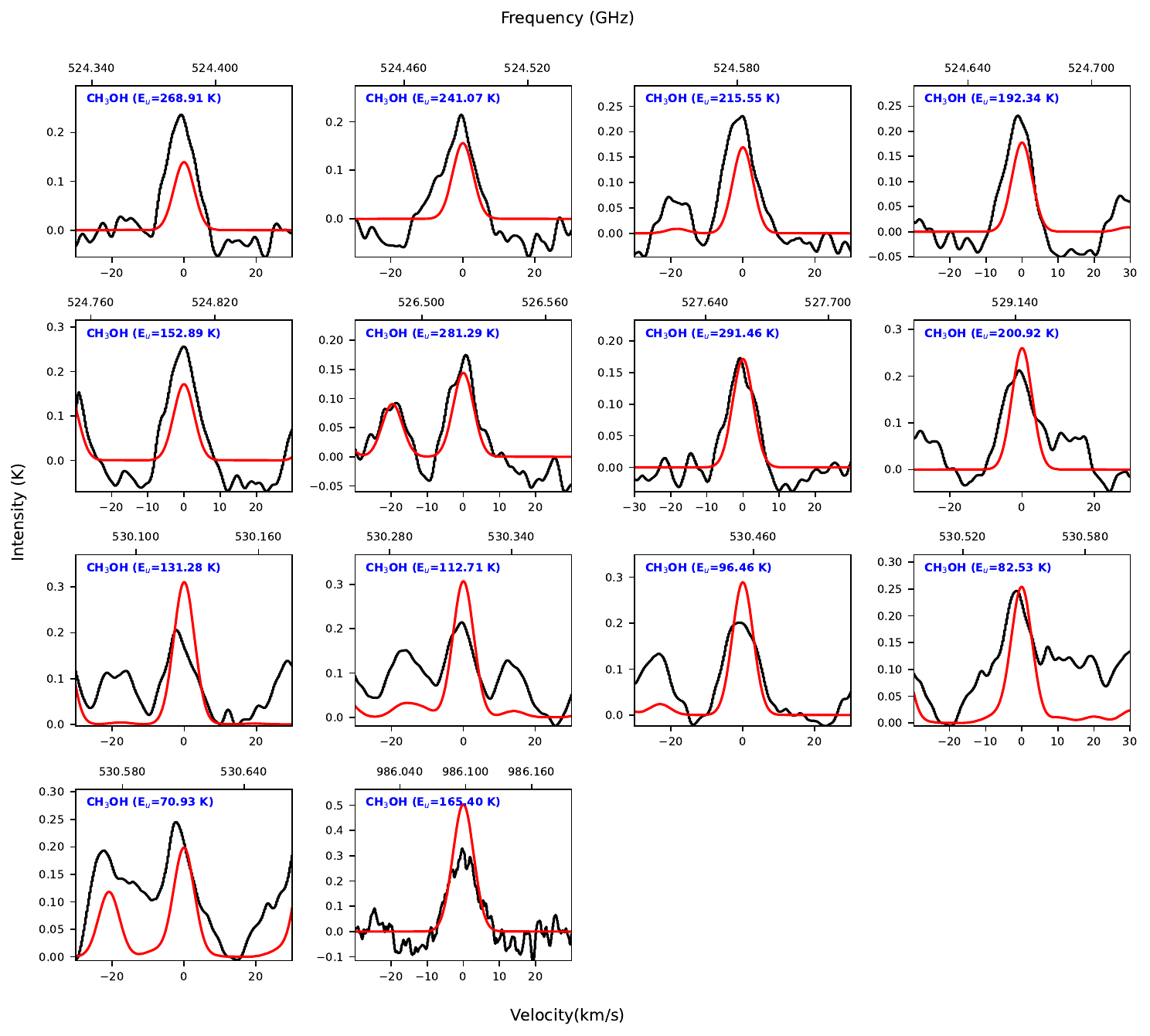}\\
    \caption{MCMC ﬁtting of observed lines of CH$_{3}$OH (vt=0-2) below 300 K toward G10. Black lines represent the observed spectra and red lines the synthetic spectra.}
    \label{fig:MCMC_CH3OH_1}
\end{figure}

\begin{figure}[h]
    \centering
	\includegraphics[width=\linewidth]{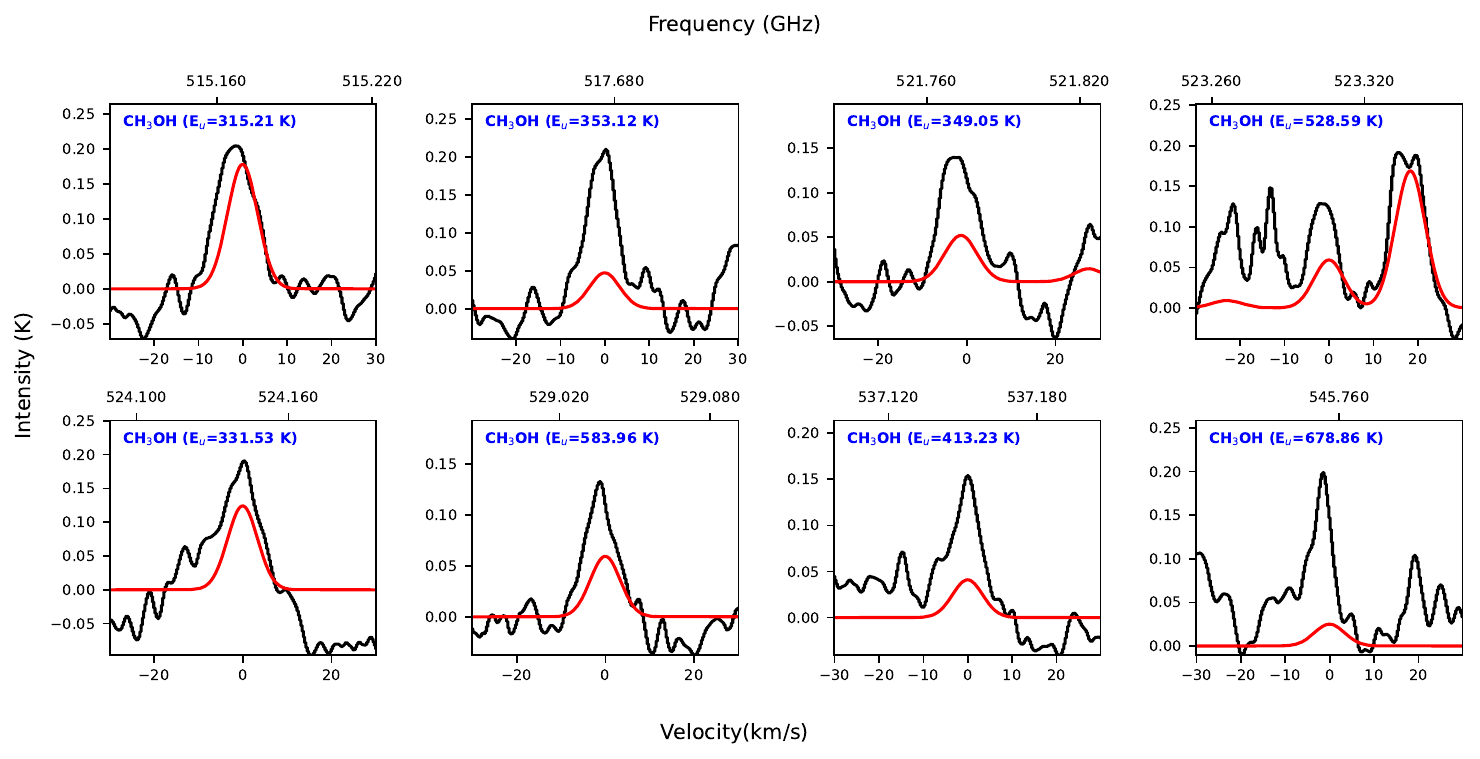}\\
    \caption{MCMC ﬁtting of observed lines of CH$_{3}$OH (vt=0-2) above 300 K toward G10. Black lines represent the observed spectra and red lines the synthetic spectra.}
    \label{fig:MCMC_CH3OH_2}
\end{figure}

The line parameters obtained with these calculations are summarized in Table~\ref{tab: MCMC}. The corresponding observed spectra, overlaid with the best-fit LTE models, are presented in Figures \ref{fig:MCMC_CH3OH_1}, \ref{fig:MCMC_CH3OH_2}, and \ref{fig:MCMC_Others}.

\begin{table}[h]
%\begin{sidewaystable}[h]
\centering
%\tiny
\caption{Summary of the best-fitted line parameters obtained by using the MCMC method.}
\vskip 0.2cm
\label{tab: MCMC}
\resizebox{\linewidth}{!}{%
\newcolumntype{H}{@{}>{\setbox0=\hbox\bgroup}c<{\egroup}@{}}

\begin{tabular}{|c|p{2 cm}|H|p{1.4 cm}|c|c|c|c|c|c|c|c|c|}
\hline
% \multirow{Species} & \multirow{Frequency (GHz)} & \multirow{Frequency Range (GHz)} & \multirow{FWHM Range (km.s$^{-1}$)} & \multirow{Best fit FWHM} & \multicolumn{2}{c|}{Column Density (cm$^{-2}$)} & \multicolumn{2}{c|}{$T_{\mathrm{ex}}$ (K)} & \multirow{Source Size ($''$)} & \multicolumn{2}{c|}{$V_{\mathrm{LSR}}$ (km.s$^{-1}$)} & \multirow{Optical Depth ($\tau$)} \\
Species & Frequency (GHz) & Frequency Range (GHz) & FWHM Range (km.s$^{-1}$) & Best fit FWHM & \multicolumn{2}{c|}{Column Density (cm$^{-2}$)} & \multicolumn{2}{c|}{$T_{\mathrm{ex}}$ (K)} & Source Size ($''$) & \multicolumn{2}{c|}{$V_{\mathrm{LSR}}$ (km.s$^{-1}$)} & Optical Depth ($\tau$) \\
\cline{6-9} \cline{11-12}
 & & & & & Range & Best fit & Range & Best fit & & Range & Best fit & \\
\hline
&517.3543&&&&&&&&&&&$1.34 \times 10^{-2}$\\
SO & 516.3354 & 515--517 & 6--12 & $6.99 \pm 0.45$ & $1.0 \times 10^{13}$--$1.0 \times 10^{15}$ & $(3.60 \pm 0.98) \times 10^{14}$ & 20--100 & $45.72 \pm 5.55$ & $39.45 \pm 0.77$ & 66--68 & $66.97 \pm 0.71$ & $1.76 \times 10^{-2}$\\
\hline
SO$_2$ & 529.9749 & 529.5--530.5 & 6--10 & $9.90 \pm 0.08$ & $1.0 \times 10^{13}$--$1.0 \times 10^{15}$ & $(2.90 \pm 0.87) \times 10^{14}$ & 20--100 & $31.66 \pm 6.70$ & $39.32 \pm 0.26$ & 66.5--67.5 & $67.44 \pm 0.04$ & $1.70 \times 10^{-2}$ \\
\hline
\multicolumn{13}{|c|}{component 1}\\
CS& 538.68883 & 538--539 & 4--7 & $5.1 \pm 0.1$ & $1.0 \times 10^{12}$--$2.0 \times 10^{16}$ & $(7.85 \pm 5.4) \times 10^{13}$ & 20--60 & $31.0 \pm 1.1$ & $34.3 \pm 0.4$ & 64--67 & $65.1 \pm 0.1$ & $4.1 \times 10^{-2}$ \\
\multicolumn{13}{|c|}{component 2}\\
& & 538--539 & 8--15 & $10.6 \pm 0.2$ & $1.0 \times 10^{12}$--$2.0 \times 10^{16}$ & $(4.31 \pm 1.29) \times 10^{13}$ & 80--150 & $96.1 \pm 0.7$ & $19.8 \pm 0.37$ & 69--75 & $73.8 \pm 0.1$ & $1.0 \times 10^{-2}$ \\
\hline

C$^{34}$S & 530.0712 & 529--531 & 6--9 & $7.69 \pm 0.91$ & $1.0 \times 10^{12}$--$1.0 \times 10^{14}$ & $(2.91 \pm 0.52) \times 10^{13}$ & 20--100 & $42.49 \pm 2.13$ & $39.59 \pm 0.07$ & 64--67 & $66.86 \pm 0.18$ & $1.06 \times 10^{-2}$ \\
\hline
H$_2$$^{18}$O & 994.675129 &  & 7--10 & $9.2 \pm 0.6$  & $5.0 \times 10^{12}- 5.0 \times 10^{16}$  & $(1.37 \pm 0.5) \times 10^{14}$  & 20-70  &$23.2\pm 2.1$  & $29.2 \pm 3.4$ & 67-70 & $69.5 \pm 0.1$ & $1.4 \times 10^{-1}$ \\
\hline
H$_2$$^{17}$O & 991.520129 &  & 3--8 & $6.6 \pm 0.5$  & $1.0 \times 10^{12}- 1.0 \times 10^{15}$  & $(8.4 \pm 4.2) \times 10^{13}$  & 20-70  &$29.4\pm 2.2$  & $31.1 \pm 1.3$ & 64-69 & $66.5 \pm 1.1$ & $1.1 \times 10^{-1}$\\
\hline
&1153.126822&&&&&&&&&&&$ 8.32 \times 10^{-1}$\\
o-H$_2$O & 1162.9116 & 987.92681162--1164 & 6--8 & $7.59 \pm 0.48$ & $5.0 \times 10^{14}$--$5.0 \times 10^{16}$ & $(1.69 \pm 0.43) \times 10^{16}$ & 20--100 & $38.28 \pm 1.42$ & $18.12 \pm 0.12$ & 66--67 & $66.43 \pm 0.24$ & $5.24 \times 10^{-1}$ \\
\hline
\multicolumn{13}{|c|}{component 1}\\
p-H$_2$O & 987.9268 & 985--990 & 4--7 & $4.5 \pm 0.1$ & $1.0 \times 10^{11}$--$2.0 \times 10^{16}$ & $(6.33 \pm 0.4) \times 10^{13}$ & 20--60 & $30.0 \pm 1.1$ & $34.4 \pm 0.4$ & 62--65 & $63.5 \pm 0.1$ & $3.0 \times 10^{-1}$ \\
\multicolumn{13}{|c|}{component 2}\\
& & 985--990 & 5--15 & $11.02 \pm 0.28$ & $1.0 \times 10^{11}$--$8.0 \times 10^{16}$ & $(4.81 \pm 0.4) \times 10^{13}$ & 80--150 & $97.25 \pm 0.1$ & $22.3 \pm 0.1$ & 67--71 & $69.9 \pm 0.1$ & $3.24 \times 10^{-2}$ \\
\hline
\multicolumn{13}{|c|}{component 1}\\
HCN & 531.7164 & 530--533 & 4--7 & $5.06 \pm 0.1$ & $1.0 \times 10^{12}$--$2.0 \times 10^{16}$ & $(1.10 \pm 0.90)\times 10^{14}$ & 20--60 & $23.9 \pm 3.1$ & $36.3\pm 1.0$ & 60--66 & $63.5 \pm 0.3$ & $1.0 \times 10^{-1}$ \\
\multicolumn{13}{|c|}{component 2}\\
& & 530--533 & 5--15 & $11.9 \pm 0.4$ & $1.0 \times 10^{12}$--$5.0 \times 10^{16}$ & $(1.72 \pm 0.60) \times 10^{13}$ & 80--150 & $94.9 \pm 2.3$ & $20.3 \pm 0.6$ & 70--75 & $73.5 \pm 0.8$ & $1.52 \times 10^{-2}$ \\

\hline
HNC & 543.8974 & 542--545 & 4--7 & $5.30 \pm 0.22$ & $1.0 \times 10^{12}$--$1.0 \times 10^{16}$ & $(2.86\pm 0.70) \times 10^{13}$ & 10--100 & $21.74 \pm 1.46$ & $30.03 \pm 1.14$ & 62--67 & $63.9 \pm 0.24$ & $4.87 \times 10^{-2}$ \\
\hline
NH$_2$ & 993.322971&  & 6--8 & $7.1 \pm 0.5$  & $1.0 \times 10^{13}- 1.0 \times 10^{15}$  & $(3.6 \pm 0.6) \times 10^{14}$  & 10-100  &$30.8\pm 0.7$  & $34.1 \pm 1.9$ & 66-70 & $67.8 \pm 0.9$ & $7.3 \times 10^{-3}$ \\

\hline
HCO$^+$ & 535.0614 & 534--536 & 4--8 & $6.81 \pm 1.14$ & $1.0 \times 10^{12}$--$1.0 \times 10^{14}$ & $(3.75 \pm 1.31) \times 10^{13}$ & 10--100 & $23.29 \pm 3.82$ & $39.26 \pm 0.14$ & 64--66 & $65.34 \pm 0.38$ & $2.11 \times 10^{-1}$ \\
\hline
C$^{18}$O & 987.564 & 990--995 & 7-12  & $9.6 \pm 0.1$ & $3.0 \times 10^{13}- 8.0 \times 10^{16}$  & $(1.8 \pm 0.53) \times 10^{16}$  &10-130  & $64.5 \pm 7.9$  & $23.08 \pm 3.28$  & 66-69  & $66.7 \pm 0.1$ & $1.9 \times 10^{-2}$ \\
\hline
&991.3293&&&&&&&&&&&$1.67 \times 10^{-1}$ \\
$^{13}$CO & 1211.3297 & 990--995 & 5--10 & $7.59 \pm 1.42$ & $1.0 \times 10^{13}$--$1.0 \times 10^{18}$ & $(1.80 \pm 0.85) \times 10^{17}$ & 10--100 & $40.34 \pm 9.16$ & $18.67 \pm 0.27$ & 64--69 & $65.63 \pm 0.94$ & $1.98 \times 10^{-2}$\\
\hline
\multicolumn{13}{|c|}{Component 1}\\
CO&1036.9124&&&&&&&&&&&$2.34 \times 10^{-1}$\\
& 1151.9855 & 1035--1155 & 4--7 & $5.1 \pm 0.3$ & $1.0 \times 10^{13}$--$2.0 \times 10^{17}$ & $(5.5 \pm 2.3) \times 10^{16}$ & 20--60 & $41.3 \pm 9.9$ & $37.5 \pm 2.0$ & 62--65 & $63.3 \pm 0.2$ & $1.13 \times 10^{-1}$ \\
\multicolumn{13}{|c|}{Component 2}\\
&1036.9124&&&&&&&&&&&$5.2 \times 10^{-2}$\\
& 1151.9855 & 1035--1155 & 5--12 & $11.0 \pm 0.4$ & $1.0 \times 10^{13}$--$8.0 \times 10^{17}$ & $(3.7 \pm 0.6) \times 10^{16}$ & 80--150 & $104.5 \pm 6.5$ & $20.0 \pm 1.0$ & 67--71.5 & $70.1 \pm 0.4$ & $4.1 \times 10^{-2}$ \\
\hline
&516.3372&&&&&&&&&&&$ 2.38 \times 10^{-2}$\\
H$_{2}$CS&522.4031&516--523&6--10 & $8.37\pm 1.20$ & $1.0 \times 10^{13}$--$1.0 \times 10^{16}$ & $(1.65 \pm 0.94) \times 10^{15}$ & 20--100 & $54.52 \pm 12.95$ & $31.0 \pm 4.71$ & 65--67 & $66.38 \pm 0.2$ &$ 2.61 \times 10^{-2}$\\
\hline
&524.384667&&&&&&&&&&&$ 2.62 \times 10^{-3}$\\
&524.488918&&&&&&&&&&&$ 2.97 \times 10^{-3}$\\
&524.582604&&&&&&&&&&&$ 3.27 \times 10^{-3}$\\
&524.666219&&&&&&&&&&&$ 3.47 \times 10^{-3}$\\
&524.804977&&&&&&&&&&&$ 3.42 \times 10^{-3}$\\
&526.520131&&&&&&&&&&&$ 2.68 \times 10^{-3}$\\
&527.658026&&&&&&&&&&&$ 3.16 \times 10^{-3}$\\
&529.143188&&&&&&&&&&&$ 5.02 \times 10^{-3}$\\
CH$_3$OH (vt=0-2) &530.123296& 515--545 & 4--12 & $6.17 \pm 0.91$ & $1.0 \times 10^{13}$--$1.0 \times 10^{15}$ & $(6.24 \pm 3.31) \times 10^{14}$ & 20--200 & $117.55 \pm 23.58$ & $28.42 \pm 7.44$ & 62--67 & $66.18 \pm 0.81$ & $ 6.13 \times 10^{-3}$\\
(E$_u<$300 K)&530.316194&&&&&&&&&&&$ 6.08 \times 10^{-3}$\\
&530.454670&&&&&&&&&&&$ 5.74 \times 10^{-3}$\\
&530.549228&&&&&&&&&&&$ 5.04 \times 10^{-3}$\\
&530.610267&&&&&&&&&&&$ 3.95 \times 10^{-3}$\\
&986.097819&&&&&&&&&&&$ 6.65 \times 10^{-3}$\\
\hline
&515.170245&&&&&&&&&&&$4.16 \times 10^{-3}$\\
&516.922004&&&&&&&&&&&$1.32 \times 10^{-4}$\\
&517.676592&&&&&&&&&&&$1.09 \times 10^{-3}$\\
&521.776019&&&&&&&&&&&$6.41 \times 10^{-4}$\\
&522.04636&&&&&&&&&&&$4.35 \times 10^{-4}$\\
CH$_3$OH (vt=0-2)&523.306240& 515--545 & 4--12 & $8.58 \pm 0.47$ & $1.0 \times 10^{13}$--$1.0 \times 10^{16}$ & $(2.80 \pm 0.71) \times 10^{15}$ & 20--400 & $240.78 \pm 97.12$ & $29.74 \pm 5.14$ & 62--67 & $66.50 \pm 0.35$ &$1.35 \times 10^{-3}$\\
(E$_u>$300 K)&524.142239&&&&&&&&&&&$9.84 \times 10^{-3}$\\
&529.038126&&&&&&&&&&&$1.33 \times 10^{-3}$\\
&531.437209&&&&&&&&&&&$4.54 \times 10^{-4}$\\
&531.460073&&&&&&&&&&&$4.58 \times 10^{-4}$\\
&537.151571&&&&&&&&&&&$9.08 \times 10^{-4}$\\
&545.755816&&&&&&&&&&&$5.38 \times 10^{-4}$\\
\hline
\end{tabular}}
\end{table}

\textsc{Methanol (CH$_3$OH):} It is one of the most abundant COMs in the ISM and plays a central role in the chemical evolution of star-forming regions. Our Herschel/HIFI spectral line investigation of the hot molecular core G10 revealed multiple transitions of methanol, spanning a wide range of upper-state energy levels from 32.86 K to 783.40 K (see Table~\ref{tab:line_parameters}). 
The rotational diagram of methanol shown in Figure \ref{fig: RD_E-CH3OH} clearly indicates that better fitting results would be obtained if we consider two temperature components.
Two panels of Figure \ref{fig: RD_E-CH3OH} show that a more accurate estimate of the column density and rotational temperature for CH$_3$OH can be achieved by fitting the transitions separately based on their upper-state energies. Specifically, transitions with upper state energy below 300 K yield a rotational temperature of $132.41\pm19.02$ K with a column density of $(6.18\pm 0.48) \times 10^{14}$ cm$^{-2}$, while those above 300 K result in a rotational temperature of $227.51\pm66.76 $ K with a column density of $(1.94\pm0.35) \times 10^{15}$ cm$^{-2}$ (see Figure \ref{fig: RD_E-CH3OH}).

Applying a similar approach in the MCMC fitting process, we determined an excitation temperature of $117.55\pm 23.58$ K for transitions with energy levels below 300 K, corresponding to a column density of $(6.24\pm 3.31) \times 10^{14}$ cm$^{-2}$. For transitions with energy levels above 300 K, the excitation temperature was found to be $240.78\pm97.12$ K, with a column density of $(2.80\pm 0.71) \times 10^{15}$ cm\(^{-2}\) (see Table \ref{tab: MCMC}). These findings align well with the rotational diagram.

These excitation circumstances are compatible with the hot core environment, where dust temperatures above the sublimation threshold of ice mantles ($\sim$100 K) cause thermal desorption of methanol and other molecules into the gas phase \citep{Garrod2008}. The low optical depths $(\tau < 0.1)$ indicate a narrow emission. The extensive excitation conditions indicate that methanol spreads throughout compact, warm, and long moderate-temperature areas, similar to what is reported in other hot cores, such as Orion KL and Sgr B2(N) \citep{Crockett14, Tahani16}.

\textsc{Carbon monoxide (CO)} and its \textsc{isotopologues $^{13}$CO and C$^{18}$O}: Carbon monoxide (CO) is the most abundant molecule after H$_2$ and serves as a key tracer of molecular gas in star-forming regions. Toward G10, we detect the $J = 9{-}8$ and $J = 10{-}9$ transitions of CO, with upper-state energies ($E_u$) of 249~K and 304~K, respectively. 
The CO line profiles clearly exhibit two distinct components with respect to the systemic velocity (67 km s$^{-1}$): a narrow blueshifted component and a wing profile in the redshifted zone. We modeled these profiles using a two-component MCMC fit, and the derived parameters are summarized in Table \ref{tab: MCMC}. 
The two-component fit indicates that the narrow, blueshifted component, located 3.5 km s$^{-1}$ from the systemic velocity, has a higher column density but a lower temperature. This component may primarily represent the extended outflow. In contrast, the broad, redshifted component, found at 3 km s$^{-1}$ from the systemic velocity, is characterized by a lower column density and a higher excitation temperature. This component likely corresponds to a minor fraction of the gas heated and kinematically disturbed, possibly due to protostellar activity or outflow-driven shocks near the core.

In addition to the high excitation transition of CO, we also identify high excitation transitions of $^{13}$CO as $J=9{-}8$ and  $J=11{-}10$ having E$_u= 237.93$~K and 348.92~K respectively, as well as a transition for C$^{18}$O ($9{-}8$) having E$_u= 237.03$~K. Unlike carbon monoxide, we only observe a single peak for these isotopologues. 

The peaks of $^{13}$CO and C$^{18}$O correspond to the systematic velocity of the source. The measured excitation temperatures are $40.34\pm 9.16$ K and $64.5\pm 7.9$ K, respectively, which are equal or greater than the blue-shifted narrow component ($41.3\pm 9.9$ K) obtained for CO but lower than the red-shifted component ($104.5\pm 6.5$ K) in Table \ref{tab: MCMC}. Therefore, these transitions indicate both the quiescent envelope and the outflow component.

Table \ref{tab:line_parameters} shows that the integrated intensity ratio of $^{12}$CO to $^{13}$CO is 2.7. 
Similarly, the intensity ratio of $^{12}$CO to C$^{18}$O is 7.5. 
These ratios are substantially lower than the canonical local-ISM value
($^{12}$C/$^{13}$C isotopic ratio of 57–74, \cite{lang93}, $^{16}$O/$^{18}$O ratio of approximately 500, \cite{lodders20094}). This
likely reflects an optical-depth effect, with $^{12}$CO appearing optically thick. We have estimated the $^{12}$CO line integrated intensity based on that obtained from its rare isotopologues, $^{13}$CO (considering $^{12}$CO/$^{13}$CO ~ 65 ), which yields a $^{12}$CO/C$^{18}$O ratio of 178.

\textsc{Water (H$_2$O):} High excitation transitions of ortho-water (o-H$_2$O) and para-water (p-H$_2$O) have been detected toward G10 and noted in Table \ref{tab:line_parameters}. 
Interestingly, the line profile of p-H$_2$O exhibits a behavior similar to that of CO. Accordingly, we performed a two-component fit, and the resulting line parameters are listed in Table \ref{tab: MCMC}. The narrow blueshifted component (shifted by 3.5 km s$^{-1}$ from the systematic velocity) has an FWHM of $4.5\pm 0.1$ km s$^{-1}$ for p-H$_2$O, compared to $5.1\pm0.3$ km s$^{-1}$ for CO, with derived excitation temperatures of $30.0\pm1.1$ K and $41.3\pm 9.9$ K, respectively. For the broad redshifted component (shifted by ~3 km s$^{-1}$ from the systematic velocity), both CO and p-H$_2$O show a FWHM of $\sim$11 km s$^{-1}$, with excitation temperatures of $104.5\pm6.5$ K for CO and $97.25\pm 0.1$ K for p-H$_2$O. We also detected isotopologues of H$_2$O: H$_2$$^{18}$O and H$_2$$^{17}$O, with an upper state energy of 100.6 K. The best-fit column densities are noted in Table \ref{tab: MCMC}.

\textsc{Hydrogen cyanide (HCN)} and \textsc{hydrogen isocyanide (HNC):} These are well-known indicators of dense molecular gas in star-forming regions \citep{Hacar20}. Like CO and p-H$_2$O, the HCN line profile also exhibits a similar pattern, with both blueshifted and redshifted components clearly visible. We fitted the data using two components, and the resulting parameters are presented in Table \ref{tab: MCMC}. Similar to the findings for CO and p-H$_2$O, we observed that the narrow blueshifted component has a higher column density and a lower temperature compared to the redshifted component, which has a lower column density and a higher temperature. Notably, for HCN, the redshifted component is significantly shifted from the systematic velocity by 6.5 km s$^{-1}$, which is different from the values observed for CO and p-H$_2$O. This suggests that the emission might originate from dense shocked clumps or the base of an outflow lobe, which is moving at a comparatively faster velocity than the region where CO and p-H$_2$O emissions were observed. Thus, HCN might trace a localized shocked region, while CO and p-H$_2$O probably probe the broader warm flow.

However, confirming this requires an imaging analysis of these species, which is beyond the scope of this work. We obtained emission data for HNC, which displays a single peak centered at the systemic velocity. Consequently, we analyzed it using a single component. Interestingly, the single peak arises close to the blueshifted component of the HCN.

\textsc{Carbon monosulfide (CS)} and its \textsc{isotopologue (C$^{34}$S):} The observations of $J=11-10$ transitions of CS and C$^{34}$S toward G10 provide insights into the dense gas properties of a high-mass star-forming core \cite{Plume1997}. Like CO, p-H$_2$O, and HCN, the profile of the CS line is better represented by a two-component fit shown in Fig. \ref{fig:MCMC_Others} and Table \ref{tab: MCMC}. Our two-component fit shows a similar characteristic: a blueshifted component with a higher column density and lower temperature, in contrast to the redshifted wing, which has a lower column density and a higher temperature. Like HCN, its redshifted wing is more shifted from the systemic velocity than CO and p-H$_2$O. 
We detected emission from the isotopologue C$^{34}$S, which exhibits a single peak at the systemic velocity. A comparison between the total column density derived for CS and that of C$^{34}$S yields a ratio lower than the canonical elemental $^{32}$S/$^{34}$S value, indicating that the main CS transition is affected by optical depth and is likely optically thick.

\textsc{Formyl cation (HCO$^+$):} It is one of the most important tracers of dense molecular gas, as its abundance is sensitive to the ionization fraction, and it is strongly enhanced in shocked or UV-irradiated environments. In G10, the $6 \rightarrow 5$ transition at 535.0614~GHz exhibits a line width of $6.81\pm1.14$~km~s$^{-1}$, a column density of $(3.75 \pm 1.31) \times 10^{13}$~cm$^{-2}$ and an excitation temperature of $23.29 \pm 3.82$~K. These results indicate moderate excitation, consistent with conditions in other high-mass star-forming cores, where HCO$^+$ effectively traces both quiescent gas and dynamically active regions associated with feedback processes \citep{Gerner2014}.

\textsc{Sulfur monoxide (SO)} and \textsc{Sulfur dioxide (SO$_2$):} Transitions of SO and SO$_2$ toward G10, trace regions affected by protostellar feedback and shocks.  The SO transitions $12_{1,2}-11_{1,1}$ and $12_{1,3}-11_{1,2}$ show a broad line width ($6.99 \pm 0.45$ km s$^{-1}$), a moderate excitation temperature ($T_{\rm ex} = 45.72 \pm 5.55$ K) and a column density of $(3.60 \pm 0.98) \times 10^{14}$ cm$^{-2}$, all of which indicate a turbulent warm gas.  Similar features are displayed by SO$_2$, which has an excitation temperature of $31.66 \pm 6.70$ K and a column density of $(2.90 \pm 0.87) \times 10^{14}$ cm$^{-2}$.  According to \cite{Bachiller1997, Esplugues14}, these sulfur-bearing species are known to act as shock tracers and are frequently amplified in areas where the outflows disturb the surrounding medium, releasing molecules from the mantles of the dust particles. In G10, their presence facilitates dynamic chemical processing and active high-mass star formation.

\textsc{Thioformaldehyde (H$_2$CS):} The transitions of H$_2$CS in G10 exhibit a column density of $(1.65 \pm 0.94) \times 10^{15}$ cm$^{-2}$ with an excitation temperature of $54.52 \pm 12.95$ K. The moderately broad line widths $8.37\pm 1.20$ km s$^{-1}$ shown in Table \ref{tab: MCMC} and the extended emission suggest its association with warm, dense, and possibly shocked gas.  A valuable indicator of early-stage sulfur chemistry in hot cores, H$_2$CS is known to arise by gas-phase and grain-surface processes \citep{Hatchell1998, vanderTak2003}.  Its existence confirms the high temperatures and vigorous chemical evolution in high-mass star-forming regions.

\textsc{Amino radical (NH$_2$):} The hyperfine transitions of the amino radical at 993.322971~GHz ($E_{\mathrm{u}} = 167.8$~K) are observed toward G10. We obtain a good fit to the observed spectra with a column density of $(3.6 \pm 0.6) \times 10^{14}$~cm$^{-2}$ and an excitation temperature of 31~K.

Here, CO, p-H$_2$O, HCN, and CS exhibit a similar profile.
The last column of Table \ref{tab: MCMC} shows the optical depth for each transition under the LTE approximation, indicating that all considered transitions are optically thin. To further verify this, we conducted a separate analysis using the RADEX code, a non-LTE radiative transfer tool for modeling molecular line emissions \citep{vandertak2007}. For our analysis, we adopted a hydrogen number density of \(8.52 \times 10^5\) cm\(^{-3}\), derived from the parameters fitted to our spectral energy distribution (SED) (see Section \ref{sec:sed}). Additionally, we assumed the excitation temperature, full width at half maximum (FWHM), and column density values listed in Table \ref{tab: MCMC}. For each species with available collisional rates, we utilized data from the LAMDA database \citep{scho05}, except for C$^{34}$S. For H$_2$CS, collisional rates for the observed transitions were not available. Under these conditions, we confirmed that all transitions investigated were optically thin. Also in Table \ref{tab: Abundances}, estimated abundances of the species were derived from the column density obtained through the MCMC fitting.

 In addition to the detected emission lines, we identify two transitions of OH$^+$ and two transitions of NH in absorption, along with one absorption transition of H$_2$S and one transition of ammonia. Previous observations by \cite{vand21} have reported OH$^+$ in absorption toward G10.47+0.03. A detailed radiative transfer and chemical modeling of these absorption features is beyond the scope of the present work and will be presented in a forthcoming study.

\tiny
\begin{table}[h]
\tiny
\centering
\caption{Estimated abundances were derived from the column density obtained through the MCMC fitting presented in Table \ref{tab: MCMC}. For this estimation, we consider $N_{\mathrm{H_2}} = 6.74 \times 10^{23}$ $\mathrm{cm^{-2}}$, as indicated in Section \ref{sec:sed}. \label{tab: Abundances}}
\vskip 0.5cm
\begin{tabular}{c c c}		
\hline								
Species  &Column Density (cm$^{-2}$) & Fractional Abundance\\		
\hline
SO & $(3.60 \pm 0.98) \times 10^{14}$ & $(5.34 \pm 3.94)\times 10^{-10}$  \\
\hline
SO$_2$ &  $(2.90 \pm 0.87) \times 10^{14}$ & $(4.30 \pm 3.33)\times 10^{-10}$ \\
\hline
CS & $(7.85 \pm 5.4) \times 10^{13}$ & $(1.16 \pm 1.37)\times 10^{-10}$\\
& $(4.31 \pm 1.29) \times 10^{13}$ & $(0.64 \pm 0.49)\times 10^{-10}$\\
\hline
C$^{34}$S & $(2.91 \pm 0.52) \times 10^{13}$ & $(0.43 \pm 0.26)\times 10^{-10}$ \\
\hline
H$_2^{18}$O &$(1.37 \pm 0.5) \times 10^{14}$ & $(2.03 \pm 1.74)\times 10^{-10}$ \\
\hline
H$_2^{17}$O &$(8.40 \pm 4.20) \times 10^{13}$ & $( 1.25\pm1.24 )\times 10^{-10}$ \\
\hline
o-H$_2$O & $(1.69 \pm 0.43) \times 10^{16}$ & $(2.51 \pm 1.79)\times 10^{-8}$ \\
\hline
p-H$_2$O & $(6.33 \pm 0.4) \times 10^{13}$ & $(0.94 \pm 0.33)\times 10^{-10}$ \\
& $(4.81 \pm 0.4) \times 10^{13}$ & $(0.71 \pm 0.29)\times 10^{-10}$\\
\hline
HCN & $(1.10 \pm 0.90) \times 10^{14}$ & $(1.63 \pm 2.09)\times 10^{-10}$ \\
& $(1.72 \pm 0.60) \times 10^{13}$ & $(2.55 \pm 2.13)\times 10^{-11}$ \\
\hline
HNC &  $(2.86 \pm 0.70) \times 10^{13}$ & $(4.24 \pm 2.96)\times 10^{-11}$\\
\hline
NH$_2$ &  $(3.60 \pm 0.60) \times 10^{13}$ & $( 5.34\pm3.08 )\times 10^{-11}$\\
\hline
HCO$^+$ & $(3.75 \pm 1.31) \times 10^{13}$ & $(5.56 \pm 4.65)\times 10^{-11}$ \\
\hline
C$^{18}$O  & $(1.8 \pm 0.53) \times 10^{16}$ &  $(2.67 \pm 2.05)\times 10^{-8}$\\
\hline
$^{13}$CO & $(1.80 \pm 0.85) \times 10^{17}$ &  $( 2.67\pm2.59 ) \times 10^{-7}$\\
\hline
CO& $(5.5 \pm 2.3) \times 10^{16}$ &  $(8.16 \pm 7.46) \times 10^{-8}$\\

& $(3.7 \pm 0.6) \times 10^{16}$ & $(5.49 \pm 3.13) \times 10^{-8}$\\
\hline
H$_{2}$CS& $(1.65 \pm 0.94) \times 10^{15}$ & $(2.44 \pm 2.61) \times 10^{-9}$ \\
\hline
%A-CH$_3$OH & $(6.57 \pm 1.59) \times 10^{14}$ & $(0.97 \pm 0.68) \times 10^{-9}$\\
%\hline
CH$_3$OH (E$_u<300$~K)& $(6.24 \pm 3.31) \times 10^{14}$ & $( 9.26\pm 9.53) \times 10^{-10}$\\
\hline
CH$_3$OH (E$_u>300$~K)& $(2.80 \pm 0.71) \times 10^{15}$ & $(4.15 \pm 2.96 ) \times 10^{-9}$\\
\hline
\end{tabular}\\
 \end{table}
% %%%%%%%%%%%%%%%%%%%%%%%%%%%%%%%%%%%%%%%%%%%%%%%%%%%%%%%%%%%%%%%%%%%%%%%%%%%%%%%%%%%%%%%%%%%%
\subsection{New Insights from the Herschel High-Frequency Observations}

Earlier observations establish G10 as a chemically rich, compact hot molecular core embedded in a massive clump and closely associated with ultracompact H\,\textsc{ii} regions \citep{cesa10,rolf11,Gorai2020,Mondal21,mondal2023}, but a uniform far-IR/submm line census or a detailed excitation and kinematic decomposition across many key tracers has not been reported. Against this backdrop, the new contribution of the present \textit{HIFI} archival data analysis is a homogeneous far-IR spectral view of this source.  This study expands the inventory of detected species to include multiple isotopologues and high-excitation ($>$100~K) transitions of CO, HCN, $^{13}$CO, C$^{18}$O, CS, C$^{34}$S, SO, ortho-/para-H$_2$O, H$_2^{18}$O, H$_2$CS, and CH$_3$OH. In particular, the detection of a high-excitation CO transition in this source is new and directly probes warmer and denser gas than the low-$J$ CO transitions commonly targeted in earlier work, extending the diagnostic reach of CO to the hot inner regions sampled by \textit{HIFI}. The CH$_3$OH rotational-diagram analysis reveals two temperature components ($\sim$132~K from low-$E_{\rm up}$ lines and $\sim$228~K from high-$E_{\rm up}$ lines), supported by MCMC fitting. This result places the methanol-emitting gas into at least two physically distinct temperature regimes, refining earlier single-temperature characterizations often assumed when only limited transitions spanning a narrow range of $E_{\rm up}$ were available. Moreover, the simultaneous detection of grain-surface products (CH$_3$OH and ortho-/para-H$_2$O, including H$_2^{18}$O) together with multiple sulfur-bearing species (SO, SO$_2$, and CS isotopologues) provides a chemically consistent picture in which shocks or outflow activity desorb ice mantles and rapidly drive sulfur chemistry, thereby linking the chemistry more directly. Finally, by fitting line profiles of CO, para-H$_2$O, HCN, and CS with distinct narrow and broad components, the analysis separates cooler, narrow outflow-layer gas from warmer, dynamically active gas likely associated with inner outflows or shocks. This complements the prior view of G10 as a compact, active hot core by adding a quantitative, multi-tracer kinematic decomposition in the far-IR. Furthermore, various protostellar properties have been derived through spectral energy distribution (SED) analysis using archival \textit{Spitzer}/IRAC (3.6--8.0~$\mu$m), MIPS 24~$\mu$m, and \textit{Herschel}/PACS--SPIRE (70--500~$\mu$m) data. This analysis also provides estimates of column densities, which are subsequently used to determine the fractional abundances of all detected species.

\section{Conclusions \label{sec:conclusions}}
In this work, we present a spectral survey of G10 using Herschel archival data covering the frequency range 515–1226 GHz. We provide a detailed analysis of its molecular line inventory, which reveals a chemically rich and dynamically complex environment typical of high-mass star-forming regions.This study represents the first investigation of the spectral properties of this source over this frequency range. The main conclusions are as follows. 

$\bullet$ We detected a wide range of molecular species shown in Table \ref{tab: Species}, including CO, HCN, HCO$^+$, $^{13}$CO, C$^{18}$O, HNC, CS, C$^{34}$S,
SO, SO$_2$, o-H$_2$O, p-H$_2$O, H$_2$$^{18}$O, H$_2$CS, and CH$_3$OH. Multiple transitions of CH$_3$OH and H$_2$CS
were identified.
This is the first time that a high excitation transition of carbon monoxide (CO) has been identified in this source.

$\bullet$ The rotational diagram of CH$_3$OH in Figure \ref{fig: RD_E-CH3OH} reveals two temperature components. The first component, with
$T=132.41\pm19.02$ K is derived from low-excitation transitions ($E_\mathrm{up} < 300$ K)
of CH$_3$OH. The second, warmer component, with $T=227.51\pm 66.76$ K, is obtained from the high-excitation
transitions ($E_\mathrm{up} > 300$ K) of CH$_3$OH. Our MCMC results in Table \ref{tab: MCMC} further support this. Our measured temperatures from MCMC fitting are consistent with those obtained using the rotational diagram analysis.

$\bullet$  We detect sulfur-bearing species, methanol, and water in both ortho and para forms.  
Methanol and water are known to form predominantly on dust-grain ice mantles. The coexistence of these species with sulfur-bearing molecules-commonly enhanced in shocked environments—strongly suggests shock-induced desorption of ice mantles followed by rapid gas-phase chemical processing.
Isotopologue detections across multiple excitation conditions further constrain the thermal structure of the core and its density distribution. 

$\bullet$ Our best-fit SED modeling constrains the key physical parameters of G10, including the visual extinction, the total mass of the dense molecular core surrounding the protostar, the mass of the central protostar, the mass surface density of the core, the envelope mass, the core radius, and the bolometric luminosity estimated both under the assumption of isotropic emission and as an average over different inclination angles. These constraints provide crucial insight into the evolutionary stage of G10, helping to refine our understanding of the physical processes driving massive star formation in this region. They also serve as a benchmark for future molecular line studies and detailed chemo-dynamical modeling.

$\bullet$  The line profiles obtained for CO, p-H$_2$O, HCN, and CS exhibit similarities. The MCMC fitting indicates that the narrow components likely trace the outer layer of the outflow. This is indicated by their obtained best fit with small velocity widths ($\sim 5$ km s$^{-1}$) and blue-shifted peaks (shifted by $\sim$3.5 km s$^{-1}$) relative to the systemic velocity, along with a temperature range of 24 K to 41 K. In contrast, the broad red-shifted (shifted by $\sim3$ km s$^{-1}$) for CO and p-H$_2$O and by 6.5-6.8 km s$^{-1}$ for HCN and CS probably originates from dynamically active gas associated with the inner layers of the outflows or shocks, which has a temperature of approximately 100 K. 
While the line survey provides detailed chemical and kinematic information, complementary imaging observations would offer spatial context and reveal the distribution of these species across the source.

\section*{Data Availability Statement}
The original contributions presented in the study are included in the article/Supplementary Material; further inquiries can be directed to the corresponding author.
\section*{Conflict of Interest Statement}
The authors declare that the research was conducted in the absence of any commercial or financial relationships that could be construed as a potential conflict of interest.
\section*{Author Contributions}
All authors listed have made a substantial, direct, and intellectual contribution to the work and approved it for publication.
\section*{Acknowledgment}
R.F. acknowledges support from the grants PID2023-146295NB-I00, and from the Severo Ochoa grant CEX2021-001131-S funded by MCIN/AEI/ 10.13039/501100011033 and by ``European Union NextGenerationEU/PRTR''.
\clearpage

\begin{figure}[h]
    \centering
	\includegraphics[width=\linewidth]{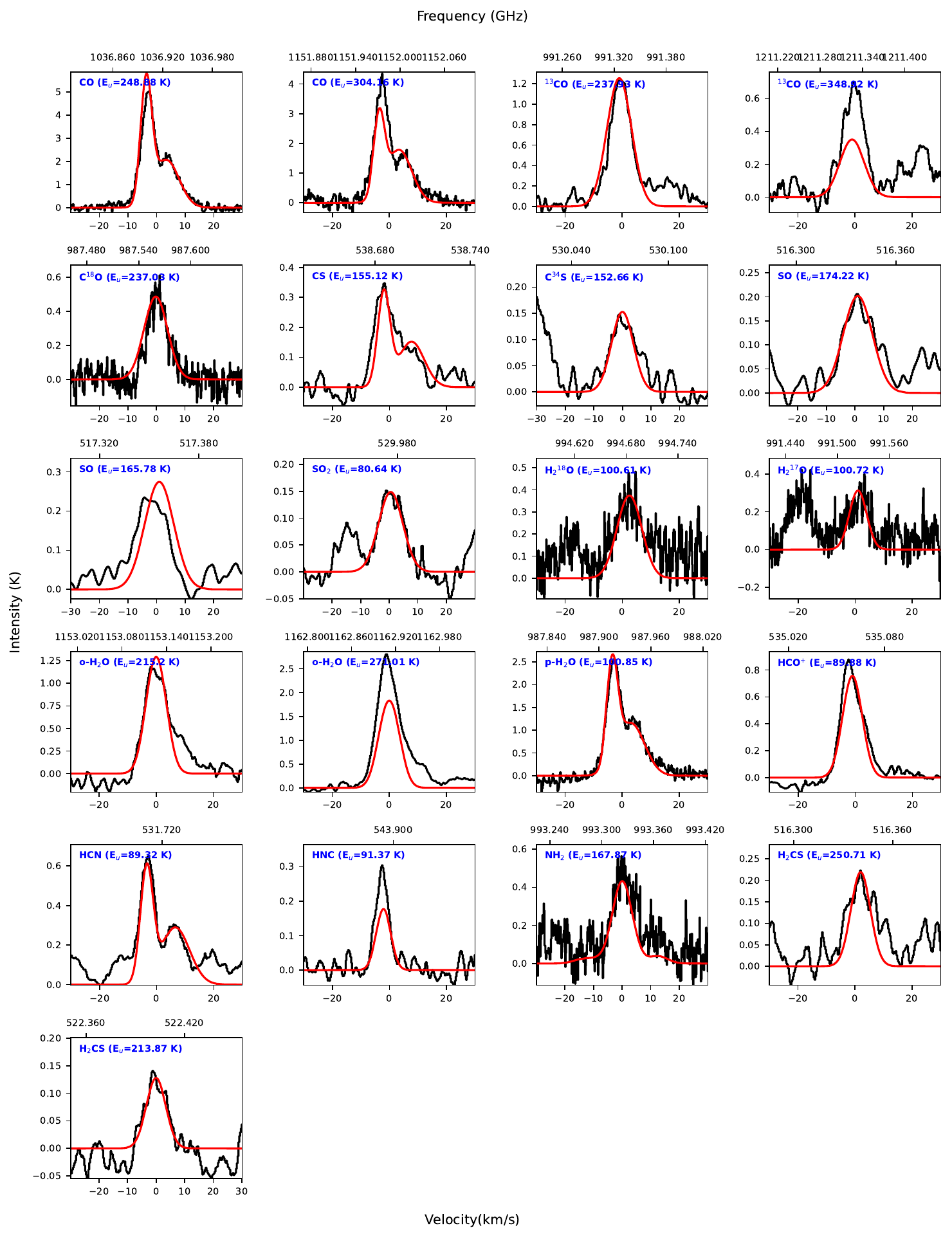}\\
    \caption{MCMC ﬁtting of observed lines of CO and its isotopologues, CS, C$^{34}$S, SO, SO$_2$, H$_2$O, HCO$^{+}$, HCN, HNC, NH$_2$, and H$_2$CS  toward G10. Black lines
represent the observed spectra and red lines the synthetic spectra.}
    \label{fig:MCMC_Others}
\end{figure}

\clearpage
%\bibliographystyle{frontiersinSCNS_ENG_HUMS}
% \bibliographystyle{Frontiers-Harvard} 
% \bibliography{reference}

    %\bibliography{reference.bbl}
    
\end{document}